\documentclass[aps,draft,twocolumn,floatfix]{revtex4}

\usepackage{bm}

\input epsf

\begin{document}

\title{
Exact wave functions for an electron on a graphene triangular quantum dot
}
\author{
A.V. Rozhkov$^{1, 2}$, 
and
Franco Nori$^{\rm 2,3}$
}

\affiliation{
$^{1}$ Institute for Theoretical and Applied Electrodynamics Russian
Academy of Sciences, 125412, Moscow, Russia
}
\affiliation{
$^{2}$  Advanced Science Institute, The Institute of Physical and Chemical
Research (RIKEN), Wako-shi, Saitama, 351-0198, Japan 
}
\affiliation{
$^{3}$ Department of Physics,
The University of Michigan, Ann Arbor, MI 48109-1040, USA
}
%\maketitle

\begin{abstract}
We generalize the known solution of the Schr\"odinger equation, describing
a particle confined to a triangular area, for a triangular graphene quantum
dot with armchair-type boundaries. The quantization conditions, wave
functions, and the eigenenergies are determined analytically. As an
application, we calculate the corrections to the quantum dot's energy
levels due to distortions of the carbon-carbon bonds at the edges of the
quantum dot.
\end{abstract}

\date{\today}

\maketitle
\hfill
%\draft

\section{Introduction}

Graphene attracts considerable attention due to its unusual electronic
properties, including: large mean free path, ``relativistic" dispersion of
the low-lying electron states, and ``valley" degeneracy (see, e.g., reviews
\cite{neto_etal,cresti,geim}).
These remarkable features suggest that some day graphene mesoscopic
structures might revolutionize nanoscience. Thus, a substantial amount of
effort has been invested studying graphene nanodevices, such as quantum
dots (QDs)
\cite{dot_reference},
bilayer structures
\cite{2-layer},
nanoribbons
\cite{nanoribbon_reference,fujita,louieI,gunlycke,rozh_sav_nori},
and other objects: e.g, $p-n$ junctions, superlattices (both magnetic and
non-magnetic), and samples with gates
\cite{gate_slatt}.

In this paper we study graphene QDs, which are interesting and important
nanodevices. A significant characteristic of a QD is its single-electron
spectrum; that is, its single-electron wave functions and corresponding
eigenenergies. There is substantial body of literature dedicated to
investigating the single-electron spectral properties of graphene QDs using
numerical tools (see, e.g., Refs.~\cite{qd_numerics,akola,akola_PRB}).
Instead of using numerical approaches, in this paper we obtain an
{\it analytical} solution for a QD shaped as an equilateral triangle
(triangular QD, or TQD) with armchair-type edges. (Some analytical results
for a TQD with zigzag edges are reported in
\cite{hawrylak}.)

The basis of our construction is the {\it exact} solution of the wave
equation inside an area shaped like an equilateral triangle. This solution
is considered for different contexts in \cite{solution}.  The present study
is inspired by Ref.~\cite{akola,akola_PRB}, where the results of
Refs.~\cite{solution} were used as a tool of analysis for TQD
single-electron {\it numerical} data. 

Once the wave functions are found, to demonstrate their usefulness, we
derive corrections to the single-electron levels of the TQD due to
deformation of the carbon-carbon bonds at the edges of the dot.

This paper is organized as follows. In
Sect.~\ref{well}
we describe the solution of the Schr\"odinger equation inside a triangular
well. The necessary basic graphene physics is outlined in
Sect.~\ref{graphene}.
In
Sect.~\ref{graphene_dot}
analytical expressions for the single-electron wave functions for a
graphene triangular quantum dot are found. The properties of these wave
functions are investigated in
Sect.~\ref{properties}.
The corrections due to the edge bond deformation are calculated in
Sect.~\ref{deformations}.
The obtained results are discussed in
Sect.~\ref{discussion}.

\section{A quantum particle inside a triangular well}
\label{well}

Since in this paper we study a triangular graphene QD, as a preparatory
discussion, let us derive the wave function for a quantum particle, confined
inside an infinitely deep triangular well. Investigating such a system we
avoid complications the graphene lattice introduces to the problem, yet the
most salient features of the wave function are brought to light. Thus, we
want to solve the Schr\"odinger equation:
\begin{eqnarray}
E \psi(x,y)
=
\frac{p^2}{2M} \psi(x,y),
%%%%%%%%%%%%%%%%%%%%%%%%%%%%%%%%%%%
\label{schr}%%%%%%%%%%%%%%%%%%%%%%%
%%%%%%%%%%%%%%%%%%%%%%%%%%%%%%%%%%%
\end{eqnarray}
with the wave function 
$\psi(x,y)$
vanishing at the boundaries of the equilateral triangle with side $L$:
\begin{eqnarray}
\psi(x,0) = 0,
%%%%%%%%%%%%%%%%%%%%%%%%%%%%%%%%%%
\label{bc1}%%%%%%%%%%%%%%%%%%%%%%%
%%%%%%%%%%%%%%%%%%%%%%%%%%%%%%%%%%
\\
\psi(x,\sqrt{3} x) = 0,
%%%%%%%%%%%%%%%%%%%%%%%%%%%%%%%%%%
\label{bc2}%%%%%%%%%%%%%%%%%%%%%%%
%%%%%%%%%%%%%%%%%%%%%%%%%%%%%%%%%%
\\
\psi(x,\sqrt{3}L - \sqrt{3}x) = 0.
%%%%%%%%%%%%%%%%%%%%%%%%%%%%%%%%%%
\label{bc3}%%%%%%%%%%%%%%%%%%%%%%%
%%%%%%%%%%%%%%%%%%%%%%%%%%%%%%%%%%
\end{eqnarray}
Equation (\ref{schr}) and the boundary conditions Eqs.(\ref{bc1}-\ref{bc3})
constitute a well-defined eigenvalue problem.

As a preliminary step for solving this problem, let us ignore
Eq.~(\ref{bc3}) for the time being and construct a wave function, which
satisfies Eqs.~(\ref{bc1}) and (\ref{bc2}). This amounts to solving
Eq.~(\ref{schr}) inside an infinite sector limited by the lines $y=0$ and
$y=\sqrt{3}x$.

To find the wave function inside the sector we imagine that there is an
incoming plane wave with wave vector
${\bf k}_1$:
\begin{eqnarray}
\psi_1 = \exp( - i {\bf k}_1 {\bf r} ),
\\
{\bf k}_1 = (k_x, k_y),
%%%%%%%%%%%%%%%%%%%%%%%%%%%%%%%%%
\label{k1}%%%%%%%%%%%%%%%%%%%%%%%
%%%%%%%%%%%%%%%%%%%%%%%%%%%%%%%%%
\\
k_y \ne \pm \sqrt{3} k_x,\ k_y \ne 0.
%%%%%%%%%%%%%%%%%%%%%%%%%%%%%%%%%%
\label{k_limitations}%%%%%%%%%%%%%
%%%%%%%%%%%%%%%%%%%%%%%%%%%%%%%%%%
\end{eqnarray}
The boundary $y=0$ reflects this wave into another plane wave 
\begin{eqnarray}
\psi_2 = \exp( - i {\bf k}_2 {\bf r} ),
\end{eqnarray}
with wave vector
\begin{eqnarray}
{\bf k}_2 = (k_x, -k_y).
\end{eqnarray}
Now the difference
$(\psi_1 - \psi_2)$
satisfies Eq.~(\ref{bc1}).
These two plane waves are reflected by the boundary
$y=\sqrt{3}x$,
creating two additional plane waves 
$\psi_{5,6}$,
whose wave vectors are:
\begin{eqnarray}
{\bf k}_5 = - \frac{1}{2} \left(
				k_x + \sqrt{3} k_y,
				-\sqrt{3} k_x + k_y
			  \right),
%%%%%%%%%%%%%%%%%%%%%%%%%%%%%%%%%
\label{k5}%%%%%%%%%%%%%%%%%%%%%%%
%%%%%%%%%%%%%%%%%%%%%%%%%%%%%%%%%
\\
{\bf k}_6 = - \frac{1}{2} \left(
				k_x - \sqrt{3} k_y,
				-\sqrt{3} k_x - k_y
			  \right).
\end{eqnarray}
These two also experience a reflection at the
$y=0$
boundary, inducing two additional plane waves
$\psi_{3,4}$
with
\begin{eqnarray}
{\bf k}_3 = - \frac{1}{2} \left(
				k_x - \sqrt{3} k_y,
				\sqrt{3} k_x + k_y
			  \right),
%%%%%%%%%%%%%%%%%%%%%%%%%%%%%%%%%
\label{k3}%%%%%%%%%%%%%%%%%%%%%%%
%%%%%%%%%%%%%%%%%%%%%%%%%%%%%%%%%
\\
{\bf k}_4 = - \frac{1}{2} \left(
				k_x + \sqrt{3} k_y,
				\sqrt{3} k_x - k_y
			  \right).
%%%%%%%%%%%%%%%%%%%%%%%%%%%%%%%%%
\label{k4}%%%%%%%%%%%%%%%%%%%%%%%
%%%%%%%%%%%%%%%%%%%%%%%%%%%%%%%%%
\end{eqnarray}
\begin{figure}[btp]
\centering
\leavevmode
\epsfxsize=8.5cm
\epsfbox{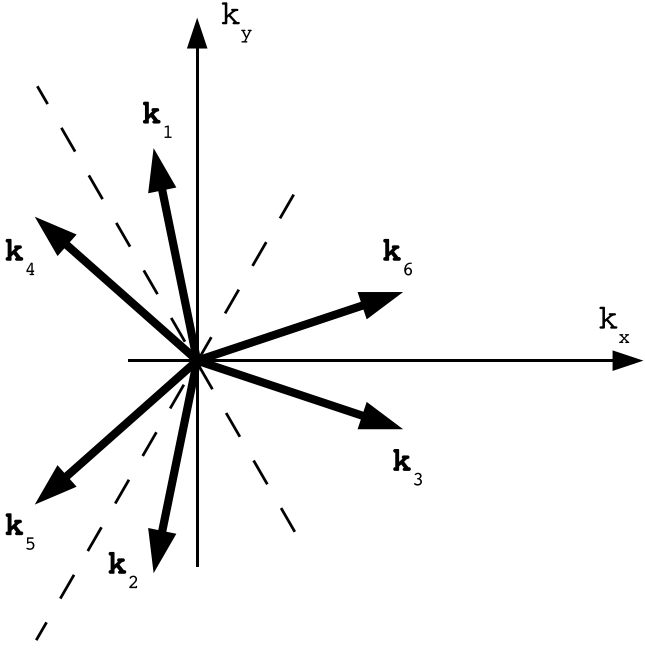}
\caption[]
{\label{sextet}
Sextet of plane waves which compose the wave function
$\psi$
in Eq.~(\ref{psi}). The dashed lines are 
$k_y = \pm \sqrt{3} k_x$.
}
\end{figure}
Fortunately, when these two undergo reflection at the
$y=\sqrt{3}x$
boundary, no new plane wave appears. The sextet of wave vectors
(Fig.~\ref{sextet})
is closed under reflections with respect to the sector's boundaries. A set
of six plane waves
$\psi_\alpha = \exp(-i{\bf k}_\alpha {\bf r})$,
$\alpha = 1\ldots 6$,
is enough to describe the wave function inside the sector. The wave
function in question is:
\begin{eqnarray}
\psi &=& \psi_1 - \psi_2 +
\psi_3 - \psi_4 +
\psi_5 - \psi_6
%%%%%%%%%%%%%%%%%%%%%%%%%%%%%%%%%%%
\label{psi}%%%%%%%%%%%%%%%%%%%%%%%%
%%%%%%%%%%%%%%%%%%%%%%%%%%%%%%%%%%%
\\
\nonumber 
&=&
\sum_{\alpha = 1}^6
(-1)^{\alpha+1} \psi_\alpha.
\end{eqnarray}
By construction it satisfies the boundary conditions Eq.~(\ref{bc1}) and
Eq.~(\ref{bc2}), see Fig.~\ref{wfunc}.
%%%%%%%%%%%%%%%%%%%%%%%%%%%%%%%%%%%%%%%%%%%%%%%%%%%
\begin{figure}[btp]
\centering
\leavevmode
\epsfxsize=8.5cm
\epsfbox{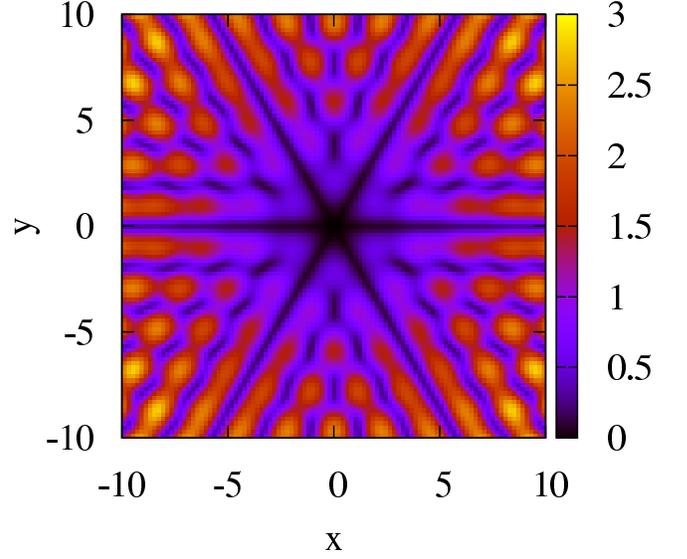}
\caption[]
{\label{wfunc}
(Color online.) The absolute value
$|\psi(x,y)|$
of the wave function
$\psi(x,y)$,
Eq.~(\ref{psi}), for arbitrary ${\bf k}$. The wave function vanishes at 
$y=0$
and at
$y=\pm \sqrt{3} x$.
}
\end{figure}
%%%%%%%%%%%%%%%%%%%%%%%%%%%%%%%%%%%%%%%%%%%%%%%%%%%

Finally, we need to enforce the third boundary condition, Eq.~(\ref{bc3}).
On the line
\begin{eqnarray}
{\bf r} = {\bf r}_0 + {\bf v} s, {\rm where\ } 
\\
{\bf r}_0 = (L,0),
\ 
{\bf v} = \frac{1}{2}(-1, \sqrt{3}),
%%%%%%%%%%%%%%%%%%%%%%%%%%%%%%%%%%%%%
\label{r0_v}%%%%%%%%%%%%%%%%%%%%%%%%%
%%%%%%%%%%%%%%%%%%%%%%%%%%%%%%%%%%%%%
\end{eqnarray} 
and $s$ varies from zero to $L$, our wave function is equal to:
\begin{eqnarray}
\psi ({\bf r}_0 + {\bf v} s)
=
\Big\{
\left[
	\exp(-i{\bf k}_1 {\bf r}) - \exp(-i{\bf k}_4 {\bf r})
\right]
%%%%%%%%%%%%%%%%%%%%%%%%%
\label{bc3_cond}%%%%%%%%%
%%%%%%%%%%%%%%%%%%%%%%%%%
\\
\nonumber
+
\left[
	\exp(-i{\bf k}_3 {\bf r}) - \exp(-i{\bf k}_2 {\bf r})
\right]
\\
\nonumber
+
\left[
	\exp(-i{\bf k}_5 {\bf r}) - \exp(-i{\bf k}_6 {\bf r})
\right]
\Big\}\Big|_{{\bf r} = {\bf r}_0 + {\bf v}s}.
\end{eqnarray} 
We now group the plane waves
$\exp(-i{\bf k}_1 {\bf r})$
and
$\exp(-i{\bf k}_4 {\bf r})$
together because 
${\bf v}{\bf k}_1 = {\bf v}{\bf k}_4$,
see Fig.~\ref{sextet} and definitions Eqs.~(\ref{k1}) and (\ref{k4}). For
the same reason we cluster
$\exp(-i{\bf k}_3 {\bf r})$
with
$\exp(-i{\bf k}_2 {\bf r})$,
and
$\exp(-i{\bf k}_5 {\bf r})$
is grouped with
$\exp(-i{\bf k}_6 {\bf r})$.
As a result, the value of $\psi$ on the boundary can be expressed as:
\begin{eqnarray} 
\psi ({\bf r}_0 + {\bf v} s)
=
A \exp\left[
		\frac{i}{2}
		\left(
			k_x - \sqrt{3} k_y
		\right)s
	\right]
\\
\nonumber
+
B \exp\left[
		\frac{i}{2}
		\left(
			k_x + \sqrt{3} k_y
		\right)s
	\right]
+
C \exp (-ik_x s).
\end{eqnarray}
The coefficients are:
\begin{eqnarray}
A&=&\exp (-ik_x L) - \exp[ i (k_x + \sqrt{3} k_y)L/2],
\\
B&=&\exp[ i (k_x - \sqrt{3} k_y)L/2] - \exp (-ik_x L),
\\
C&=&\exp[ i (k_x + \sqrt{3} k_y)L/2]
\\
\nonumber
&& - \exp[ i (k_x - \sqrt{3} k_y)L/2].
\end{eqnarray}
Equation (\ref{bc3_cond}) vanishes, if $A$, $B$, and $C$ are all equal to
zero. This occurs when the following conditions are met:
\begin{eqnarray}
k_x = \frac{2\pi}{3L}(n-m),
%%%%%%%%%%%%%%%%%%%%%%%%%%%%%%%%%%%%
\label{quant_cond_x}%%%%%%%%%%%%%%%%
%%%%%%%%%%%%%%%%%%%%%%%%%%%%%%%%%%%%
\\
k_y = \frac{2\pi}{\sqrt{3}L}(n+m).
%%%%%%%%%%%%%%%%%%%%%%%%%%%%%%%%%%%%
\label{quant_cond_y}%%%%%%%%%%%%%%%%
%%%%%%%%%%%%%%%%%%%%%%%%%%%%%%%%%%%%
\end{eqnarray}
Here $n$ and $m$ are integers. Equations~(\ref{quant_cond_x}) and
(\ref{quant_cond_y}) are the quantization conditions for the particle
momentum due to confinement. The wave function
$\psi^{n,m}(x,y)$
with momentum 
${\bf k}_1$
satisfying these equations is the solution of the Schr\"odinger equation
with boundary conditions Eqs.~(\ref{bc1}-\ref{bc3}) and the eigenvalue:
\begin{eqnarray}
E = \frac{8 \pi^2 \hbar^2}{9ML^2}(n^2 + m^2 + nm).
%%%%%%%%%%%%%%%%%%%%%%%%%%%%%%%%%%%
\label{Enm}%%%%%%%%%%%%%%%%%%%%%%%%
%%%%%%%%%%%%%%%%%%%%%%%%%%%%%%%%%%%
\end{eqnarray}
The wave function
$\psi^{n,m}$
vanishes identically, if any of the equalities
\begin{eqnarray}
n=0,\ 
{\rm or\ }
m=0,\ 
{\rm or\ }
n = -m,
%%%%%%%%%%%%%%%%%%%%%%%%%%%%%%%%%
\label{wf_vanish}%%%%%%%%%%%%%%%%
%%%%%%%%%%%%%%%%%%%%%%%%%%%%%%%%%
\end{eqnarray} 
holds [for example, if
$n = -m$, 
then
$k_y=0$,
$\Rightarrow$
${\bf k}_1 = {\bf k}_2$,
${\bf k}_3 = {\bf k}_6$,
${\bf k}_5 = {\bf k}_4$,
and both the even and odd terms of Eq.~(\ref{psi}) cancel each other].
Conditions
Eq.~(\ref{wf_vanish})
are equivalent to
Eq.~(\ref{k_limitations}).

In section~\ref{graphene_dot} we show how to adopt
$\psi^{n,m}$
for a graphene TQD.

\section{Basic physics of a graphene sheet}
\label{graphene}

For completeness, in this section we quickly remind the reader the basic
single-electron properties of a graphene sheet. Our treatment follows
Ref.~\cite{neto_etal}.
The notation introduced in this section will be used in the rest of the
paper.

It is common to describe a graphene sample in terms of a tight-binding model
on the honeycomb lattice. Such lattice can be split into two sublattices,
denoted by
${\cal A}$
and
${\cal B}$.
The Hamiltonian of an electron hopping on the graphene sheet is given by:
\begin{eqnarray}
H = 
-t
\sum_{{\bf R} \in {\cal A}} 
\sum_{i=1,2,3}
c^\dagger_{\bf R}
c^{\vphantom{\dagger}}_{{\bf R} + {\bm \delta}_i} 
+
{\rm H.c.},
%%%%%%%%%%%%%%%%%%%%%%%%%%%%%
\label{H}%%%%%%%%%%%%%%%%%%%%
%%%%%%%%%%%%%%%%%%%%%%%%%%%%%
\end{eqnarray}
where `H.c.' stands for `Hermitian conjugate', 
${\bf R}$
runs over sublattice
${\cal A}$:
\begin{eqnarray}
{\bf R} 
=
{\bm \delta}_1 + {\bf a}_1 n_1 + {\bf a}_2 n_2
\; \Leftrightarrow \;
{\bf R}
\in
{\cal A},
%%%%%%%%%%%%%%%%%%%%%%%%%%%%%%%%
\label{subA}%%%%%%%%%%%%%%%%%%%%
%%%%%%%%%%%%%%%%%%%%%%%%%%%%%%%%
\end{eqnarray}
where the primitive vectors of the honeycomb lattice are:
\begin{eqnarray}
{\bf a}_1 &=& a_0 (3/2, \sqrt{3}/2),
\\
{\bf a}_2 &=& a_0 (3/2, -\sqrt{3}/2),
\end{eqnarray}
and
$n_{1,2}$
are integers.
The symbol $a_0$ denotes the carbon-carbon bond length, which is about 1.4
\AA.
The vectors 
${\bm \delta}_i$
($i=1,2,3$)
connect the nearest neighbours. They are:
\begin{eqnarray}
{\bm \delta}_1 
&=&
a_0 (-1, 0),
\\
{\bm \delta}_2 
&=& 
a_0 (1/2, \sqrt{3}/2),
\\
{\bm \delta}_3 
&=& 
a_0 (1/2, -\sqrt{3}/2).
\end{eqnarray}
The corresponding Schr\"odinger equation can be written as:
\begin{eqnarray}
\varepsilon \, \psi^{\cal A}_{\bf R} 
&=&
- t \,
\psi^{\cal B}_{{\bf R} + {\bm \delta}_1}
-
t \sum_{i=1,2}
	\psi^{\cal B}_{{\bf R} + {\bm \delta}_1 + {\bf a}_i},
%%%%%%%%%%%%%%%%%%%%%%%%%%%%%
\label{sch_a}%%%%%%%%%%%%%%%%
%%%%%%%%%%%%%%%%%%%%%%%%%%%%%
\\
\varepsilon \, \psi^{\cal B}_{{\bf R} + {\bm \delta}_1} 
&=& 
- t \, \psi^{\cal A}_{\bf R} 
- 
t \sum_{i=1,2}
	\psi^{\cal A}_{{\bf R} - {\bf a}_i},
%%%%%%%%%%%%%%%%%%%%%%%%%%%%%%%
\label{sch_b}%%%%%%%%%%%%%%%%%%
%%%%%%%%%%%%%%%%%%%%%%%%%%%%%%%
\end{eqnarray}
where 
$\psi_{\bf R}^{\cal A}$ 
($\psi_{{\bf R} + {\bm \delta}_1}^{\cal B}$) 
denotes the wave function value at the site 
${\bf R}$ 
(at the site ${\bf R} + {\bm \delta}_1$)
of sublattice ${\cal A}$ (sublattice ${\cal B}$).

The primitive cell of graphene contains two atoms, one at
${\bf R}$,
another at
${\bf R} + {\bm \delta}_1$.
Therefore, it is convenient to define the two-component (spinor) wave
function:
\begin{eqnarray}
\Psi_{\bf R} 
= 
\left(
	\matrix{ 
			\psi_{\bf R}^{{\cal A} \hphantom{\delta_1}}\cr
			\psi_{{\bf R} + {\bm \delta}_1}^{\cal B}\cr
		}
\right).
%%%%%%%%%%%%%%%%%%%%%%%%%%%%%
\label{spinor}%%%%%%%%%%%%%%%
%%%%%%%%%%%%%%%%%%%%%%%%%%%%%
\end{eqnarray} 
By construction, the function
$\Psi_{\bf R}$ 
is defined on sublattice 
${\cal A}$,
Eq.~(\ref{subA}). 

The action of $H$ on a plane wave
\begin{eqnarray} 
\Psi_{\bf R} = \Psi_{\bf k} \exp ( -i {\bf k}{\bf R} )
\end{eqnarray} 
can be expressed as:
\begin{eqnarray}
H \Psi_{\bf k} 
=
\left(
	\matrix{ 
			0            &  -t_{\bf k}  \cr
       		       -t_{\bf k}^*  &        0     \cr
       		}
\right)
\Psi_{\bf k},
\\
t_{\bf k} 
=
t
\left[
	1 
	+ 
	2 {\exp}	\left(
				-i\frac{3  k_x a_0}{2} 
			\right)
	\cos \left(
			\frac{\sqrt{3}}{2} k_y a_0
	      \right)
\right].
%%%%%%%%%%%%%%%%%%%%%%%%%%%%%%%%%%%%%%%
\label{tk}%%%%%%%%%%%%%%%%%%%%%%%%%%%%%
%%%%%%%%%%%%%%%%%%%%%%%%%%%%%%%%%%%%%%%
\end{eqnarray}
For every ${\bf k}$ there are two eigenstates:
\begin{eqnarray}
\Psi_{{\bf k} \pm}
=
\left(
	\matrix{
				1			\cr
			\mp {\rm e}^{-i \theta_{\bf k}} \cr
		}
\right),
\\
\exp 	\left(
	{i \theta_{\bf k}} 
	\right)
=
\frac{t_{\bf k}}{|t_{\bf k}|},
%%%%%%%%%%%%%%%%%%%%%%%%%%%%%%%%%%%%%
\label{theta_def}%%%%%%%%%%%%%%%%%%%%
%%%%%%%%%%%%%%%%%%%%%%%%%%%%%%%%%%%%%
\end{eqnarray} 
with eigenvalues:
\begin{eqnarray}
\varepsilon_{{\bf k} \pm}
=
\pm |t_{\bf k}|
%%%%%%%%%%%%%%%%%%%%%%%%%%%%%%%%%%
\label{graphene_energy}%%%%%%%%%%%
%%%%%%%%%%%%%%%%%%%%%%%%%%%%%%%%%%
=
\pm t
\sqrt{
	3 
	+ 
	F({\bf k})
     },
\\
F({\bf k})
=
4\cos \left( 
		\frac{3}{2} k_x a_0 
	\right)
\cos \left( 
			\frac{\sqrt{3}}{2} k_y a_0 
     \right)
%%%%%%%%%%%%%%%%%%%%%%%%%%%%%%%%%%%
\label{F}%%%%%%%%%%%%%%%%%%%%%%%%%%
%%%%%%%%%%%%%%%%%%%%%%%%%%%%%%%%%%%
\\
\nonumber
+
2 \cos \left( 
		\sqrt{3} k_y a_0 
	\right).
\end{eqnarray}
The states with negative (positive) energy are filled (empty) at $T=0$.

\begin{figure}[btp]
\centering
\leavevmode
\epsfxsize=8.5cm
\epsfbox{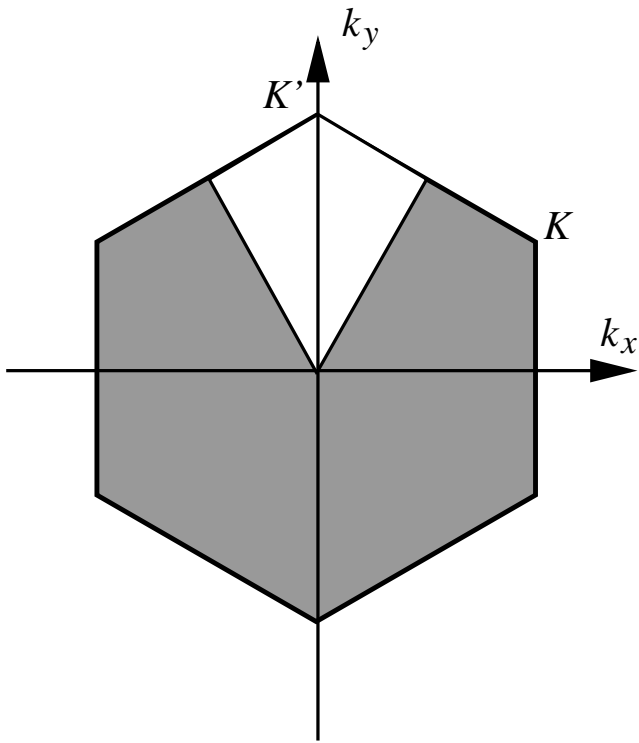}
\caption[]
{\label{bz}
The hexagon shown is the Brillouin zone of graphene. The white polygon is
where the allowed wave vectors for the triangular quantum dot are located
(see
Fig.~\ref{label} and Fig.~\ref{count}).
}
\end{figure}

The allowed values of ${\bf k}$ lie within the Brillouin zone presented on
Fig.~\ref{bz}.
%%%%%%%%%%%%%%%%%%%%%%%%%%%%%%%%%%%%%%%%%%%%%%%%%%
%%%%%%%%%%%%%%%%%%%%%%%%%%%%%%%%%%%%%%%%%%%%%%%%%%%
The reciprocal lattice is characterized by the following lattice vectors:
\begin{eqnarray}
{\bf d}_1 = (4\pi/3 a_0, 0),
%%%%%%%%%%%%%%%%%%%%%%%%%%%%%%%%%%%%%%%%
\label{d1}%%%%%%%%%%%%%%%%%%%%%%%%%%%%%%
%%%%%%%%%%%%%%%%%%%%%%%%%%%%%%%%%%%%%%%%
\\
{\bf d}_2 = ( - 2\pi/3 a_0, 2\pi/\sqrt{3} a_0 ).
%%%%%%%%%%%%%%%%%%%%%%%%%%%%%%%%%%%%%%%%
\label{d2}%%%%%%%%%%%%%%%%%%%%%%%%%%%%%%
%%%%%%%%%%%%%%%%%%%%%%%%%%%%%%%%%%%%%%%%
\end{eqnarray} 
The amplitude
$t_{\bf k}$
and energy 
$\varepsilon_{{\bf k} \pm}$
are invariant under shifts over 
${\bf d}_{1,2}$.

The quantity 
$\varepsilon_{{\bf k} \pm}$
vanishes at the six corners of the Brillouin zone:
$(0, \pm 4\pi/(3\sqrt{3}a_0))$ 
and
$(\pm 2\pi / (3 a_0), \pm 2\pi/(3\sqrt{3}a_0))$.
These are the locations of the famous Dirac cones of graphene.

\section{Schr\"odinger equation solution for an electron on a triangular
graphene dot}
\label{graphene_dot}

In this section we find the solution of Eqs.~(\ref{sch_a}) and
(\ref{sch_b}) for a graphene TQD.

\begin{figure}[btp]
\centering
\leavevmode
\epsfxsize=8.5cm
\epsfysize=7.0cm
\epsfbox{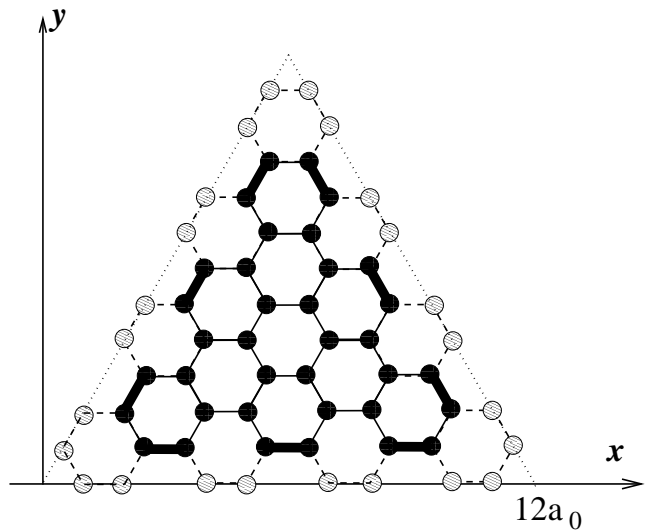}
\caption[]
{\label{tqd}
Triangular graphene quantum dot with armchair edges
(here, $N_0 = 3$ and $N=4$).
The solid lines represent covalent bonds between neighboring carbon atoms
(black circles). The thick lines at the edges represent deformed bonds,
whose effect on the spectrum is studied in
section~\ref{deformations}.
The dashed lines represent fictitious bonds connecting real carbon atoms
and auxiliary atoms. The latter are represented by hatched circles. Dotted
lines correspond to the effective boundaries of the triangular dot.
}
\end{figure}

The basic object of study here, a TQD with armchair edges, is depicted in
Fig.~\ref{tqd}. The carbon atoms are shown as black circles, the covalent
bonds are solid lines connecting the atoms. The lateral size of the TQD is
$L_0$. It is a multiple of $3a_0$:
\begin{eqnarray}
L_0 = 3 N_0 a_0,
\end{eqnarray}
where $N_0$ is an integer. The dot in Fig.~\ref{tqd} is characterized by 
$N_0 = 3$.
%%%%%%%%%%%%%%%%%%%%%%%%%%%%%%%%%%%%%%%%%%%%%%%%%%%%%%%%%%%%%%
%%%%%%%%%%%%%%%%%%%%%%%%%%%%%%%%%%%%%%%%%%%%%%%%%%%%%%%%%%%%%%

\begin{figure}[btp]
\centering
\leavevmode
\epsfxsize=8.5cm
\epsfysize=8.0cm
\epsfbox{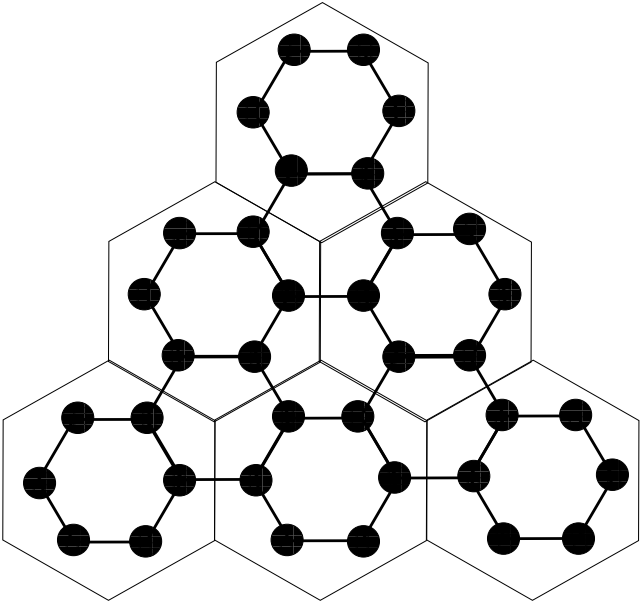}
\caption[]
{\label{dot_split}
A triangular graphene dot can be split into
$(N_0 + 1)N_0/2$
aromatic rings, where the integer
$N_0$
is proportional to
$L_0$,
the dot size:
$L_0 = 3 a_0 N_0$.
For the dot shown on the figure,
$N_0 = 3$.
Thus, the triangular dot consists of six rings.
}
\end{figure}

The total number of carbon atoms in the dot $N_a$ is:
\begin{eqnarray}
N_a = 3 N_0 (N_0 + 1).
%%%%%%%%%%%%%%%%%%%%%%%%%%%%%%%%%%%%
\label{Na}%%%%%%%%%%%%%%%%%%%%%%%%%%
%%%%%%%%%%%%%%%%%%%%%%%%%%%%%%%%%%%%
\end{eqnarray}
This formula can be derived if one splits the dot into 
$N_0(N_0+1)/2$
aromatic rings, with six atoms each (see Fig.~\ref{dot_split}).
%%%%%%%%%%%%%%%%%%%%%%%%%%%%%%%%%%%%%%%%%%%%%%%%%%%%%
%%%%%%%%%%%%%%%%%%%%%%%%%%%%%%%%%%%%%%%%%%%%%%%%%%%%%

The atoms at the edges of the dot are special for they have only two
nearest neighbors, unlike atoms in the ``bulk" of the dot, which have three
neighbors. As a result, the Schr\"odinger equations, Eq.~(\ref{sch_a}) and
Eq.~(\ref{sch_b}), for the atoms at the edges have to be modified. It is not
always convenient to work with such formalism. A simpler approach is used
in
Ref.~\cite{gunlycke,rozh_sav_nori}.
In those works it is pointed out that one may add an extra row of
carbon atoms at the armchair edges (`auxiliary' atoms, shown as hatched
circles in
Fig.~\ref{tqd})
and demand the wave function to vanish on these `atoms'. Then for a physical
(not `auxiliary') atom at the edge we do not have to amend
Eqs.~(\ref{sch_a})
and
(\ref{sch_b})
explicitly. Indeed, the absent neighbor (now represented by the `auxiliary'
atom) does not contribute to these equations, since the wave function
vanishes on the `auxiliary' atoms.

The addition of the extra row of `auxiliary' atoms slightly increases the
effective size of the dot. It is helpful to introduce the following notation:
\begin{eqnarray}
L = L_0 + 3a_0 = 3Na_0,
{\rm \ where\ }
N = N_0 + 1.
%%%%%%%%%%%%%%%%%%%%%%%%%%%%%%%%%%%
\label{tqd_L}%%%%%%%%%%%%%%%%%%%%%%
%%%%%%%%%%%%%%%%%%%%%%%%%%%%%%%%%%%
\end{eqnarray}
Although $L$ and $N$ are trivially related to $L_0$ and $N_0$, it is
convenient to define these quantities explicitly for they are heavily used
in the calculations below.

Consider now the wave function:
\begin{eqnarray}
\Psi_{{\bf R}\pm}
=
\sum_{\alpha = 1}^6 	(-1)^{\alpha + 1} \Psi_{{\bf k}_\alpha \pm}
			 	\exp (-i{\bf k}_\alpha {\bf R}),
%%%%%%%%%%%%%%%%%%%%%%%%%%%%%%
\label{tqd_wf}%%%%%%%%%%%%%%%%
%%%%%%%%%%%%%%%%%%%%%%%%%%%%%%
\end{eqnarray}
where 
${\bf k}_\alpha$
are members of a sextet. They are given in
section~\ref{well}.
Observe now that:
\begin{eqnarray}
\varepsilon_{{\bf k}_1 \pm} =
\varepsilon_{{\bf k}_2 \pm} =
\varepsilon_{{\bf k}_3 \pm} =
\varepsilon_{{\bf k}_4 \pm} =
\varepsilon_{{\bf k}_5 \pm} =
\varepsilon_{{\bf k}_6 \pm}.
%%%%%%%%%%%%%%%%%%%%%%%%%%%%%%%%%%%%
\label{sextet_eigen}%%%%%%%%%%%%%%%%
%%%%%%%%%%%%%%%%%%%%%%%%%%%%%%%%%%%%
\end{eqnarray}
This is a consequence of the graphene lattice symmetry. Therefore, the
spinor
$\Psi_{{\bf R}\pm}$
is a solution of Eqs.~(\ref{sch_a}) and (\ref{sch_b}) with eigenvalue
$\varepsilon_{{\bf k}_1 \pm}$.

Further, the upper component of 
$\Psi_{{\bf R}\pm}$
coincides with $\psi({\bf R})$, Eq.~(\ref{psi}). Thus, if ${\bf k}_1$
satisfies Eqs.~(\ref{quant_cond_x}) and (\ref{quant_cond_y}), then
$\psi^{\cal A}_{{\bf R} \pm}$
complies with the boundary condition. We remind the reader that the
zero boundary conditions must be met at the {\it effective} edges
of the TQD (on the `auxiliary' atoms).

The lower component of 
$\Psi_{{\bf R}\pm}$
requires a more tedious consideration. It is equal to:
\begin{eqnarray}
\psi^{\cal B}_{{\bf R+{\bm \delta}}_{1}\pm}
=
\mp\sum_\alpha 	(-1)^{\alpha + 1} \exp (-i\theta_{{\bf k}_\alpha})
			 	\exp (-i{\bf k}_\alpha {\bf R}).
%%%%%%%%%%%%%%%%%%%%%%%%%%%%%%%%%%
\label{psi_B1}%%%%%%%%%%%%%%%%%%%%
%%%%%%%%%%%%%%%%%%%%%%%%%%%%%%%%%%
\end{eqnarray}
As Eq.~(\ref{spinor}) specifies, the argument of
$\psi^{\cal B}$
is not ${\bf R}$, which belongs to sublattice ${\cal A}$, but rather the sum
${\bf R+{\bm \delta}}_1$,
which belongs to sublattice
${\cal B}$
(recall that
${\bf R}$
is the coordinate of the two-atom unit cell, while 
${\bf R+{\bm \delta}}_1$
is the physical location of the atom on sublattice
${\cal B}$).
We must keep this in mind when formulating the following boundary
conditions for
$\psi^{\cal B}$:
\begin{eqnarray}
\psi^{\cal B}_{{\bf R+{\bm \delta}}_{1}\pm}&=&0,
{\rm \ if\ } y = 0;
%%%%%%%%%%%%%%%%%%%%%%%%%%%%%%%%%%
\label{bb1}%%%%%%%%%%%%%%%%%%%%%%%
%%%%%%%%%%%%%%%%%%%%%%%%%%%%%%%%%%
\\
\psi^{\cal B}_{{\bf R+{\bm \delta}}_{1}\pm}&=&0,
{\rm \ if\ } {\bf R + {\bm \delta}}_1= {\bf u}s,
%%%%%%%%%%%%%%%%%%%%%%%%%%%%%%%%%%
\label{bb2}%%%%%%%%%%%%%%%%%%%%%%%
%%%%%%%%%%%%%%%%%%%%%%%%%%%%%%%%%%
\\
&&{\rm \ where\ }{\bf u} = \frac{1}{2}(1, \sqrt{3}), 
s = (3l - 1) a_0;
\\
\psi^{\cal B}_{{\bf R+{\bm \delta}}_{1}\pm}&=&0,
{\rm \ if\ } {\bf R + {\bm \delta}}_1={\bf r}_0 + {\bf v}s',
%%%%%%%%%%%%%%%%%%%%%%%%%%%%%%%%%%
\label{bb3}%%%%%%%%%%%%%%%%%%%%%%%
%%%%%%%%%%%%%%%%%%%%%%%%%%%%%%%%%%
\\
&&{\rm \ where\ } s' = (3l + 1) a_0.
\end{eqnarray}
Here $l$ is an integer;
${\bf r}_0$
and
${\bf v}$
are defined in Eq.~(\ref{r0_v}).

The first condition, Eq.~(\ref{bb1}), is fulfilled automatically. Indeed,
it is easy to check that 
$\exp(-i\theta_{\bf k})$
is independent of the sign of $k_y$. Thus
\begin{eqnarray}
\exp (-i\theta_{{\bf k}_1}) \exp (-i{\bf k}_1 {\bf R})
-
\exp (-i\theta_{{\bf k}_2}) \exp (-i{\bf k}_2 {\bf R})
\\
\nonumber
=
-2i\exp (-i\theta_{{\bf k}_1}) \exp(-ik_x x) \sin(k_y y),
\end{eqnarray}
vanishes, when $y=0$. The same holds true for the sum of the third and 
sixth terms, as well as for the sum of the fourth and fifth terms.

Let us now show that
Eq.~(\ref{bb2})
is valid. It is convenient to rewrite
Eq.~(\ref{psi_B1}) as:
\begin{eqnarray}
\psi^{\cal B}_{{\bf R+\delta}_{1}\pm}
=
\mp \sum_\alpha 
		(-1)^{\alpha + 1}
		\exp (	-i\theta_{{\bf k}_\alpha} 
			- i k_{x \alpha} a_0)
%%%%%%%%%%%%%%%%%%%%%%%%%%%%%%%%%%%%%%%%
\label{psi_B2}%%%%%%%%%%%%%%%%%%%%%%%%%%
%%%%%%%%%%%%%%%%%%%%%%%%%%%%%%%%%%%%%%%%
\\
\nonumber
\times
	 	\exp 	\left[
				-i{\bf k}_\alpha ({\bf R} + {\bm \delta}_1)
			\right].
\end{eqnarray}
When
${\bf R + {\bm \delta}}_1= {\bf u}s$,
we have:
\begin{eqnarray}
\psi^{\cal B}_{{\bf R+\delta}_{1}\pm}
=
\mp
\Big\{
	A\exp\left[
		-\frac{i}{2}(k_x + \sqrt{3}k_y)s
    	    \right]
\\
\nonumber
	+
	B\exp\left[
		-\frac{i}{2}(k_x - \sqrt{3}k_y)s
	    \right]
	+
	C\exp(ik_x s)
\Big\},
\end{eqnarray}
where
\begin{eqnarray}
A = 
\exp(-i\theta_{{\bf k}_1} - ik_{x1} a_0)
-
\exp(-i\theta_{{\bf k}_6} - ik_{x6} a_0),
\\
B = 
\exp(-i\theta_{{\bf k}_5} - ik_{x5} a_0)
-
\exp(-i\theta_{{\bf k}_2} - ik_{x2} a_0),
\\
C = 
\exp(-i\theta_{{\bf k}_3} - ik_{x3} a_0)
-
\exp(-i\theta_{{\bf k}_4} - ik_{x4} a_0).
\end{eqnarray} 
Note that the equation for $A$ involves two wave vectors: 
${\bf k}_1$
and
${\bf k}_6$.
They enter together because
${\bf u} {\bf k}_1 = {\bf u} {\bf k}_6$.
For the same reason the vectors 
${\bf k}_2$
and
${\bf k}_5$
appear in the equation for $B$, and the vectors 
${\bf k}_{3}$
and
${\bf k}_{4}$
are part of the equation for $C$. A similar structure was already observed
above, see the discussion after Eq.~(\ref{bc3_cond}).

It is easy to check that
\begin{eqnarray}
\exp(-i\theta_{{\bf k}_6})
=
\exp\left(
		- \frac{3i}{2} k_x a_0 + \frac{\sqrt{3}i}{2} k_y a_0 
    \right)
%%%%%%%%%%%%%%%%%%%%%%%%%%%%
\label{theta_rel}%%%%%%%%%%%
%%%%%%%%%%%%%%%%%%%%%%%%%%%%
\\
\nonumber 
\times
\exp(-i\theta_{{\bf k}_1}).
\end{eqnarray}
To prove this identity one has to use Eqs.~(\ref{tk}) and
(\ref{theta_def}),
and the fact that
$|t_{\bf k}| = \varepsilon_{{\bf k}+}$
is the same for all members of the sextet, see
Eq.~(\ref{sextet_eigen}).
Consequently
\begin{eqnarray}
A =
\exp( - i\theta_{{\bf k}_1} - i k_{x1} a_0)
\qquad
\qquad
%%%%%%%%%%%%%%%%%%%%%%%%%%%%%%%%%%%%%%%%%%%%%
\label{coeffA}%%%%%%%%%%%%%%%%%%%%%%%%%%%%%%%
%%%%%%%%%%%%%%%%%%%%%%%%%%%%%%%%%%%%%%%%%%%%%
\\
\nonumber
\times
\left\{
	1 -
	\exp\left[
			- ia_0
			\left(
				\frac{3}{2} k_x 
				- \frac{\sqrt{3}}{2} k_y
				- k_{x1} 
				+ k_{x6}
			\right)
	    \right]
\right\}.
\end{eqnarray}
Using the definitions of 
${\bf k}_{1}$
and
${\bf k}_{6}$
we can write:
\begin{eqnarray}
- k_{x1} + k_{x6} 
=
- \frac{3}{2} k_x +  \frac{\sqrt{3}}{2} k_y.
\end{eqnarray}
Thus, the argument of the exponential in
Eq.~(\ref{coeffA})
vanishes, and the coefficient $A$ vanishes as a result. In a similar
fashion, it is possible to prove that $B$ and $C$ are equal to zero.

Lastly, we need to demonstrate that 
$\psi^{\cal B}$ 
satisfies Eq.~(\ref{bb3}). When
${\bf R + {\bm \delta}}_1={\bf r}_0 + {\bf v}s'$,
we use Eq.~(\ref{psi_B2}) to obtain:
\begin{eqnarray}
\psi^{\cal B}_{{\bf R+\delta}_{1}\pm}
=
\mp
\Big\{
	A\exp\left[
		\frac{i}{2}(k_x - \sqrt{3}k_y)s'
    	\right]
\\
\nonumber
	+
	B\exp\left[
		\frac{i}{2}(k_x + \sqrt{3}k_y)s'
    	\right]
	+
	C\exp(-ik_x s')
\Big\},
\end{eqnarray}
with the coefficients:
\begin{eqnarray}
A&=&\exp[ -i\theta_{{\bf k}_1} - ik_x (L+a_0) ]
\\
\nonumber
   &-&
    \exp[ -i\theta_{{\bf k}_4} 
	  + 
	   \frac{i}{2}(k_x + \sqrt{3} k_y) (L+a_0)
        ],
\\
B &= &- \exp[ -i\theta_{{\bf k}_2} - ik_x (L+a_0) ]
\\
\nonumber
    &+&
    \exp[ -i\theta_{{\bf k}_3} 
	  + 
	   \frac{i}{2}(k_x - \sqrt{3} k_y) (L+a_0)
        ],
\\
C &=& \exp[ -i\theta_{{\bf k}_5} 
	  +
	   \frac{i}{2}(k_x + \sqrt{3} k_y) (L+a_0)
	  ]
\\
\nonumber
    &-&
    \exp[ -i\theta_{{\bf k}_6} 
	  + 
	   \frac{i}{2}(k_x - \sqrt{3} k_y) (L+a_0)
        ].
\end{eqnarray}
Using the relation:
\begin{eqnarray}
\exp(-i\theta_{{\bf k}_4})
=
\exp\left(
		- \frac{3i}{2} k_x a_0 - \frac{\sqrt{3}i}{2} k_y a_0 
    \right)
%%%%%%%%%%%%%%%%%%%%%%%%%%%%
\label{theta_rel2}%%%%%%%%%%
%%%%%%%%%%%%%%%%%%%%%%%%%%%%
\\
\nonumber 
\times
\exp(-i\theta_{{\bf k}_1}),
\end{eqnarray}
which is similar to Eq.~(\ref{theta_rel}) and is derived analogously,
we show that:
\begin{eqnarray}
A 
&=&
\exp[ -i\theta_{{\bf k}_1} - ik_x (L+a_0) ]
\\
\nonumber
&\times&
\left\{
	1
	-
	\exp\left[
			i L \left(
					\frac{3}{2} k_x
					+
					\frac{\sqrt{3}}{2} k_y
			    \right)
	    \right]
\right\}.
\end{eqnarray}
From here we obtain:
\begin{eqnarray}
A = 0
\; \Leftrightarrow \;
L \left(
		\frac{3}{2} k_x
		+
		\frac{\sqrt{3}}{2} k_y
    \right)
=
2\pi n.
%%%%%%%%%%%%%%%%%%%%%%%%%%%%%%%%%%%%
\label{A=0}%%%%%%%%%%%%%%%%%%%%%%%%%
%%%%%%%%%%%%%%%%%%%%%%%%%%%%%%%%%%%%
\end{eqnarray}
For the coefficient $B$, the following expression holds:
\begin{eqnarray}
B 
&=&
\exp[ -i\theta_{{\bf k}_2} - ik_x (L+a_0) ]
%%%%%%%%%%%%%%%%%%%%%%%%%%%%%%%%
\label{B_coeff}%%%%%%%%%%%%%%%%%
%%%%%%%%%%%%%%%%%%%%%%%%%%%%%%%%
\\
\nonumber
&\times&
\left\{
	-1
	+
	\exp\left[
			i L \left(
					\frac{3}{2} k_x
					-
					\frac{\sqrt{3}}{2} k_y
			    \right)
	    \right]
\right\}.
\end{eqnarray}
Deriving Eq.~(\ref{B_coeff}), we use the relation:
\begin{eqnarray}
\exp(-i\theta_{{\bf k}_3})
=
\exp\left(
		- \frac{3i}{2} k_x a_0 + \frac{\sqrt{3}i}{2} k_y a_0 
    \right)
%%%%%%%%%%%%%%%%%%%%%%%%%%%%
\label{theta_rel3}%%%%%%%%%%
%%%%%%%%%%%%%%%%%%%%%%%%%%%%
\\
\nonumber 
\times
\exp(-i\theta_{{\bf k}_2}).
\end{eqnarray}
Coefficient $B$ vanishes when
\begin{eqnarray}
L \left(
		\frac{3}{2} k_x
		-
		\frac{\sqrt{3}}{2} k_y
    \right)
=
- 2\pi m.
%%%%%%%%%%%%%%%%%%%%%%%%%%%%%%%%%
\label{B=0}%%%%%%%%%%%%%%%%%%%%%%
%%%%%%%%%%%%%%%%%%%%%%%%%%%%%%%%%
\end{eqnarray}
Coefficient $C$ vanishes automatically, when both
Eq.~(\ref{A=0})
and
Eq.~(\ref{B=0}) 
hold.

Combining
Eq.~(\ref{A=0})
and
Eq.~(\ref{B=0})
we derive the quantization condition:
\begin{eqnarray}
{\bf k}_1
=
{\bf k}^{n,m},
%%%%%%%%%%%%%%%%%%%%%%%%%%%%%%%%%%%%%%
\label{quant_cond_vec}%%%%%%%%%%%%%%%%
%%%%%%%%%%%%%%%%%%%%%%%%%%%%%%%%%%%%%%
\\
{\rm where\ }
{\bf k}^{n,m} = n {\bf K}_1 + m {\bf K}_2,
\\
{\bf K}_{1,2} 
=
\left(
	\pm \frac{2\pi}{9Na_0},
	\frac{2\pi}{3\sqrt{3}Na_0}
\right).
\end{eqnarray} 
which is equivalent to Eqs.~(\ref{quant_cond_x}) and (\ref{quant_cond_y})
with $L$ given by Eq.~(\ref{tqd_L}). The symbol
$\Psi_{\bf R}^{n,m}$
is used below to denote the wave function with
${\bf k}_1 = {\bf k}^{n, m}$.

Thus, we demonstrate that the wave function Eq.~(\ref{tqd_wf}) with
momentum quantized according to Eq.~(\ref{quant_cond_vec}) satisfies the
Schr\"odinger equations (\ref{sch_a}) and (\ref{sch_b}) on a TQD with
armchair edges. 

\section{Properties of the wave function}
\label{properties}

In this section we study the most elementary properties of the
Schr\"odinger equation solution $\Psi_{\bf R}^{n,m}$.

\subsection{Eigenenergy}
\label{eigen}

The eigenenergy corresponding to our solution is equal to:
\begin{eqnarray}
\varepsilon_{n,m \pm}&=&\varepsilon_{{\bf k}_1\pm}
=
\pm t	\left\{
		3 
		+
		2 \cos\left(
				\frac{2\pi n}{3N}
		      \right)
	\right.
%%%%%%%%%%%%%%%%%%%%%%%%%%%%%%%%%%%%
\label{eigenenergy}%%%%%%%%%%%%%%%%%
%%%%%%%%%%%%%%%%%%%%%%%%%%%%%%%%%%%%
\\
\nonumber
&&
    	\left.
		+
		2 \cos\left(
				\frac{2\pi m}{3N}
		      \right)
		+
		2 \cos\left[
				\frac{2\pi (n+m)}{3N}
		      \right]
     	\right\}^{1/2}.
\end{eqnarray}
The eigenenergy remains unchanged, if $n$ and $m$ are switched. Therefore,
if the wave functions
$\Psi^{n,m}$
and
$\Psi^{m,n}$
are linearly independent, the corresponding states are degenerate.

Let us define $\tilde n$ and $\tilde m$ as:
\begin{eqnarray}
n = N + \tilde n,
\\
m = N - \tilde m.
\end{eqnarray}
If
$N \gg 1$,
$|\tilde n| \ll N$,
and
$|\tilde m| \ll N$,
then
$
\sqrt{3N_a} \approx 3N,
$
and we may expand
Eq.~(\ref{eigenenergy})
in orders of
$\tilde n/N$
and
$\tilde m/N$:
\begin{eqnarray}
\varepsilon_{\tilde n,\tilde m\pm} \approx
\pm \frac{2\pi t}{\sqrt{3N_a}}
\sqrt{\tilde n^2 + \tilde m^2 - \tilde n \tilde m}.
\end{eqnarray}
The latter formula is reported in Refs.~\cite{akola,akola_PRB}.

An interesting phenomenon occurs in TQDs with even $N$. For such object,
consider the quantum states
$\Psi^{n,m}_{{\bf R} \pm}$
with
\begin{eqnarray}
n = \frac{3N}{2} - m.
%%%%%%%%%%%%%%%%%%%%%%%%%%%%
\label{accidental1}%%%%%%%%%
%%%%%%%%%%%%%%%%%%%%%%%%%%%%
\end{eqnarray}
In this case:
\begin{eqnarray}
\varepsilon_{n,m \pm} = \pm t,
%%%%%%%%%%%%%%%%%%%%%%%%%%%%
\label{accidental2}%%%%%%%%%
%%%%%%%%%%%%%%%%%%%%%%%%%%%%
\end{eqnarray}
independent of $m$. That is, for an even-$N$ dot the energy level at 
$\pm t$
is very degenerate. Although, here we do not investigate this feature in
detail, it seems to be an accidental degeneracy, which is lifted if one
includes longer-range hopping in the Hamiltonian.

\subsection{Symmetry of the wave function}
\label{symmetry}

The geometrical symmetry group $G$ of a TQD consists of
$\pm 2\pi/3$
rotations about the center of the dot and reflections with respect to three
bisectors. Such group is isomorphic to 
$C_{3v}$
symmetry group
\cite{Landau}.
It has two one-dimensional irreducible representations,
$A_1$
and
$A_2$;
and one two-dimensional irreducible representation $E$. 

The representation $A_1$ is trivial: it maps all the group elements on 1;
$A_2$ maps all rotations on 1 and all reflections on $-1$. The
representation $E$ maps a rotation (reflection) on a 2x2 orthogonal matrix
performing a rotation (reflection) of the two-dimensional Euclidean space.

\subsubsection{Rotation}

In order to see which eigenfunction corresponds to which representation, let
us perform a
$2\pi/3$
rotation over the center of the TQD.

\begin{figure}[btp]
\centering
\leavevmode
\epsfxsize=8.5cm
\epsfbox{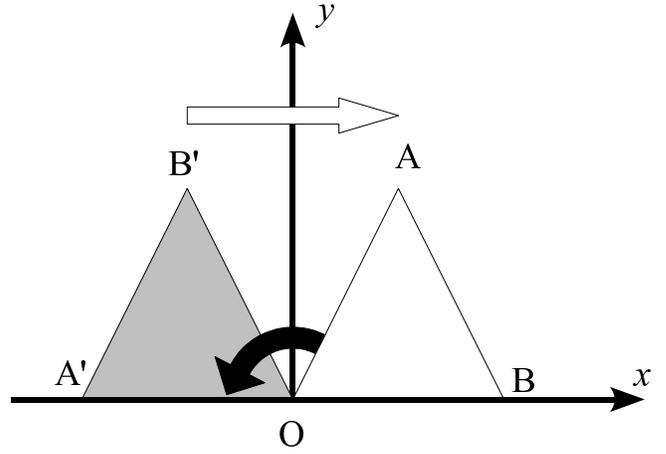}
\caption[]
{\label{rotation}
Sequence of steps to rotate the triangular quantum dot about its center.
First, the dot is rotated around the origin (black arrow). After this
transformation, the original white TQD becomes the gray TQD. Point A is
transformed into A', point B is mapped on B'. Afterwards, the dot is
shifted by $L$ (white arrow) to restore the original position.
}
\end{figure}

Technically, it is more convenient to split such transformation into two
consecutive steps: (i)
$U_{{2\pi}/{3}}$
-- a rotation about the origin over the angle 
$2\pi/3$
(such rotation does not preserve the location of the dot), followed by (ii)
a shift over $L$:
$x \rightarrow x+L$,
which restores the TQD into its position prior to step (i) (see
Fig.~\ref{rotation}).
%%%%%%%%%%%%%%%%%%%%%%%%%%%%%%%%%%%%%%%%%%%%%%%%%%%%%%%
%%%%%%%%%%%%%%%%%%%%%%%%%%%%%%%%%%%%%%%%%%%%%%%%%%%%%%%%%
After step (i) the plane wave
$\Psi_{{\bf k}_1} \exp(-i{{\bf k}_1{\bf R}})$
becomes the plane wave
$\Psi_{{\bf k}_5} \exp(-i{{\bf k}_5{\bf R}})$;
$\Psi_{{\bf k}_2} \exp(-i{{\bf k}_1{\bf R}})$
becomes the plane wave
$\Psi_{{\bf k}_6} \exp(-i{{\bf k}_5{\bf R}})$,
etc. This means that, after the rotation,
$\Psi_{\bf R}^{n,m}$
remains unchanged.

Next we perform step (ii). As a result of such shift all exponentials
acquire an extra phase factor. For example, consider
\begin{eqnarray}
\exp(-i{\bf k}_1 {\bf R}) \rightarrow
\exp(-i{\bf k}_1 {\bf R} - i k_{x1}L)
%%%%%%%%%%%%%%%%%%%%%%%%%%%%%%%%%%%%%%%%%
\label{shift}%%%%%%%%%%%%%%%%%%%%%%%%%%%%
%%%%%%%%%%%%%%%%%%%%%%%%%%%%%%%%%%%%%%%%%
\\
\nonumber
=
\exp( - i k_{x1}L) \exp(-i{\bf k}_1 {\bf R}).
\end{eqnarray}
The phase factor is:
\begin{eqnarray}
\exp( - i k_{x1}L) 
=
\exp( - i k_{x}L) 
=
\exp\left[ 
		- i \frac{2\pi}{3}(n-m)
     \right]
%%%%%%%%%%%%%%%%%%%%%%%%%%%%%%%%%%%%
\label{angular_mom}%%%%%%%%%%%%%%%%%
%%%%%%%%%%%%%%%%%%%%%%%%%%%%%%%%%%%%
\\
\nonumber
=
\cases{
	\exp\left(
			\pm \frac{2\pi i}{3}
	    \right),
	&
	if $n-m = 3p \mp 1$,
	\cr
	1,
	&
	if $n-m = 3p$,
	}
\end{eqnarray}
where $p$ is an integer. If we investigate other exponentials 
[$\exp (i {\bf k}_\alpha {\bf R})$,
$\alpha=2\ldots  6$]
we would arrive at the same expression for the phase factor. 

Thus, upon rotation around the dot's center, the wave function
$\Psi_{\bf R}^{n,m}$
acquires the phase multiplier,
Eq.~(\ref{angular_mom}).
We can say that the representation to which the wave function belongs
is fixed by the value of
$(n-m)$.
When the latter is a multiple of 3, the wave function is transformed
according to
$A_1$
or
$A_2$.
Otherwise, it is part of the two-dimensional representation $E$.

\subsubsection{Reflection}
\label{reflection}

Next, we study how 
$\Psi^{n,m}_{\bf R}$
is transformed under reflection. Another two-stage process is executed: (a)
reflection with respect to the $x=0$ line followed by (b) shift
$x \rightarrow x+L$,
which restores the original position of the TQD. This sequence reflects
the dot with respect to its vertical bisector.

After step (a) the wave function becomes
\begin{eqnarray}
\Psi_{\bf R}
\rightarrow
\left(
\matrix{ \psi^{\cal B}_{U{\bf R}} \cr
	 \psi^{\cal A}_{U{\bf R} - {\bm \delta}_1} 
       }
\right)
=
\sigma_x \Psi_{U{\bf R} - {\bm \delta}_1},
%%%%%%%%%%%%%%%%%%%%%%%%%%%%%%%%%%%%%%%%%%%
\label{reflect}%%%%%%%%%%%%%%%%%%%%%%%%%%%%
%%%%%%%%%%%%%%%%%%%%%%%%%%%%%%%%%%%%%%%%%%%
\end{eqnarray} 
where 
$\sigma_x$
is the Pauli matrix, and $U$ is the reflection transformation matrix:
$U{\bf R} = (-x, y)$.

Such a complicated transformation law is associated with the fact that the
reflection
$x \rightarrow -x$
exchanges the sublattices. Thus, the spinor components must be switched.
This is why we multiply 
$\Psi$
by
$\sigma_x$.
In addition,
$U{\bf R} \in {\cal B}$, 
while the spinor wave function should be defined on sublattice 
${\cal A}$, 
see Eqs.~(\ref{subA}) and (\ref{spinor}). Simple geometrical considerations
show that the unit cell, whose location is given by
${\bf R}$,
is reflected on the cell
$({U{\bf R} - {\bm \delta}_1})$.
Keeping the above in mind, one can derive Eq.~(\ref{reflect}). 

When
$\Psi_{\bf R}$ 
is a plane wave, Eq.~(\ref{reflect}) becomes:
\begin{eqnarray}
&&\Psi_{{\bf k} \pm} \exp(-i{{\bf k}{\bf R}})
\rightarrow
%%%%%%%%%%%%%%%%%%%%%%%%%%%%%%%%%%%%%%%%%%%%%%
\label{refl_plane}%%%%%%%%%%%%%%%%%%%%%%%%%%%%
%%%%%%%%%%%%%%%%%%%%%%%%%%%%%%%%%%%%%%%%%%%%%%
\\
\nonumber
&&\rightarrow
\mp \exp(-i \theta_{{\bf k}}+i{\bf k}{\bm \delta}_1)
\Psi_{U{\bf k}\pm} \exp(-i{{\bf k}U{\bf R}})
\\
\nonumber
&&=
\mp \exp(-i \theta_{{\bf k}} - i k_x a_0)
\Psi_{U{\bf k}\pm} \exp[-i{(U{\bf k}){\bf R}}].
\end{eqnarray}
This equation demonstrates that a plane wave with wave vector 
${\bf k}$
is mapped on a plane wave with wave vector
$U{\bf k}$,
multiplied by a phase factor
$f_{\bf k} = \exp(-i \theta_{{\bf k}} - i k_x a_0)$,
which can be expressed as:
\begin{eqnarray}
f_{\bf k}
=
\frac{t}{\varepsilon_{{\bf k}+}}
\sum_{\alpha=1}^3
	\exp \left(
			i {\bf k} {\bm \delta}_\alpha
	     \right)
=
\frac{t}{\varepsilon_{{\bf k}+}}
\sum_{s=0}^2
	\exp \left[
			i {\bf k} (U_{\frac{2\pi}{3}})^s {\bm \delta}_1
	     \right].
%%%%%%%%%%%%%%%%%%%%%%%%%%%%%%%%%%%%%%%%%%%
\label{f}%%%%%%%%%%%%%%%%%%%%%%%%%%%%%%%%%%
%%%%%%%%%%%%%%%%%%%%%%%%%%%%%%%%%%%%%%%%%%%
\end{eqnarray}
This equation may be proven with the help of Eqs.~(\ref{tk}) and
(\ref{theta_def}).

The function $f$ has three important properties:
\begin{eqnarray}
{\rm if\ }{\bf k}' = U_{\frac{2\pi}{3}} {\bf k} 
\; \Rightarrow \;
f_{{\bf k}'}
=
f_{\bf k},
%%%%%%%%%%%%%%%%%%%%%%%%%%%%%%%%%%%%%%%%%
\label{f_2pi/3}%%%%%%%%%%%%%%%%%%%%%%%%%%
%%%%%%%%%%%%%%%%%%%%%%%%%%%%%%%%%%%%%%%%%
\\
|f_{\bf k}| = 1,
\\
f_{k_x, k_y} = f_{k_x, -k_y}.
%%%%%%%%%%%%%%%%%%%%%%%%%%%%%%%%%%%%%%%%%%%
\label{f_ky}%%%%%%%%%%%%%%%%%%%%%%%%%%%%%%%
%%%%%%%%%%%%%%%%%%%%%%%%%%%%%%%%%%%%%%%%%%%
\end{eqnarray}
The first property is a simple consequence of Eq.~(\ref{f}), while the two
others follow from the definition of $f$.

Using Eqs.~(\ref{f_2pi/3}) and (\ref{f_ky}), one demonstrates that, for
all plane waves in the sextet, the phase factors
$f_{{\bf k}_\alpha }$
are identical. It is easy to prove that the transformation law for our wave
function becomes:
\begin{eqnarray}
\Psi_{{\bf R}\pm}^{n,m}
\rightarrow
\mp f_{n,m} \Psi_{{\bf R}\pm}^{m,n},
%%%%%%%%%%%%%%%%%%%%%%%%%%%%%%%%%%%%%%%%%%%%%%
\label{step_a}%%%%%%%%%%%%%%%%%%%%%%%%%%%%%%%%
%%%%%%%%%%%%%%%%%%%%%%%%%%%%%%%%%%%%%%%%%%%%%%
\\
f_{n,m} = f_{{\bf k}^{n,m}}
=
\frac{t}{\varepsilon_{n,m +}}
	\exp \left[
			- \frac{2\pi i}{9N} (n-m)
	     \right]
\\
\nonumber
\times
\left[
	1
	+
	\exp \left(
			  \frac{2\pi i}{3N} n
	     \right)
	+
	\exp \left(
			- \frac{2\pi i}{3N} m
	     \right)
\right].
\end{eqnarray}
Equation~(\ref{step_a}) shows how our wave function is transformed after
step (a) of our two-step process. 

Note that the wave function
$\Psi^{m,n}$
from the right-hand side of Eq.~(\ref{step_a}) transforms as:
\begin{eqnarray}
\Psi_{{\bf R}\pm}^{m,n}
\rightarrow
\mp f_{m,n} \Psi_{{\bf R}\pm}^{n,m}
=
\mp f_{n,m}^* \Psi_{{\bf R}\pm}^{n,m}.
%%%%%%%%%%%%%%%%%%%%%%%%%%%%%%%%%%%%%%%%%
\label{step_a'}%%%%%%%%%%%%%%%%%%%%%%%%%%
%%%%%%%%%%%%%%%%%%%%%%%%%%%%%%%%%%%%%%%%%
\end{eqnarray}
Therefore, 
$\Psi_{{\bf R}\pm}^{n,m}$,
subjected to two reflection transformations, remains unchanged, as it
should be.

The step (b) is identical to step (ii), see Eq.~(\ref{shift}) and
Eq.~(\ref{angular_mom}).
Consequently, when the TQD is subjected to the reflection about its
bisector, the wave function transforms as follows:
\begin{eqnarray}
\Psi_{{\bf R}\pm}^{n,m}
\rightarrow
	\mp f_{n,m} \exp \left[
				\frac{2 \pi i (n-m)}{3}
			 \right]
	\Psi_{{\bf R}\pm}^{m,n}.
%%%%%%%%%%%%%%%%%%%%%%%%%%%%%%%%%%%%
\label{wf_refl}%%%%%%%%%%%%%%%%%%%%%
%%%%%%%%%%%%%%%%%%%%%%%%%%%%%%%%%%%%
\end{eqnarray}

\subsubsection{One-dimensional irreducible representations $A_1$ and $A_2$}
\label{A1vsA2}

At this point we can explicitly construct the wave functions corresponding
to the representations $A_1$ and $A_2$. Recall that a wave function belongs
to a one-dimensional representation
($A_{1}$
or
$A_{2}$)
only when
$n-m = 3p$.
Assuming this relation, consider the sum:
\begin{eqnarray}
\Psi_{{\bf R}\pm \sigma }^{n, m}
=
\Psi_{{\bf R}\pm}^{n, m}  + \sigma f_{n,m} \Psi_{{\bf R}\pm}^{m, n},
%%%%%%%%%%%%%%%%%%%%%%%%%%%%%
\label{A12}%%%%%%%%%%%%%%%%%%
%%%%%%%%%%%%%%%%%%%%%%%%%%%%%
\end{eqnarray}
where 
$\sigma = \pm 1$.
Upon reflection, this wave function transforms as
[see Eq.~(\ref{wf_refl})]:
\begin{eqnarray}
\Psi_{{\bf R}\pm \sigma }^{n, m}
\rightarrow
\mp f_{n,m} \Psi_{{\bf R}\pm}^{m, n}  
\mp
\sigma \Psi_{{\bf R}\pm}^{n, m}
=
\mp \sigma \Psi_{{\bf R}\pm \sigma }^{n, m}.
%%%%%%%%%%%%%%%%%%%%%%%%%%%%%%
\label{refl}%%%%%%%%%%%%%%%%%%
%%%%%%%%%%%%%%%%%%%%%%%%%%%%%%
\end{eqnarray}
Therefore:
\begin{eqnarray}
A_1:\quad
\cases{
	\Psi_{{\bf R}+\sigma }^{n, m},& if $\sigma=- 1$, \cr
	\Psi_{{\bf R}-\sigma }^{n, m},& if $\sigma= 1$.
      }
\\
A_2:\quad
\cases{
	\Psi_{{\bf R}+\sigma }^{n, m},& if $\sigma= 1$, \cr
	\Psi_{{\bf R}-\sigma }^{n, m},& if $\sigma= - 1$.
      }
\end{eqnarray}
Since both
$\Psi_{{\bf R}\pm }^{n, m}$
and
$\Psi_{{\bf R}\pm }^{m, n}$
have identical eigenenergies
$\varepsilon_{n,m \pm}$,
their linear combination 
$\Psi_{{\bf R}\pm \sigma }^{n, m}$
also corresponds to 
$\varepsilon_{n,m \pm}$.

\subsection{Normalization of the wave function}
\label{normalization}

In order to calculate matrix elements with the help of our wave function, it
has to be normalized. Namely, it is necessary to find the coefficient 
${\gamma}$
such that:
\begin{eqnarray}
{\gamma}^2
\left(
	\sum_{{\bf R} \in {\rm TQD}}
		|\psi^{\cal A}_{\bf R}|^2 
		+
	\sum_{{\bf R}+{\bm \delta}_1 \in {\rm TQD}}
		|\psi^{\cal B}_{{\bf R} + {\bm \delta}_1}|^2  
\right)
=
1,
%%%%%%%%%%%%%%%%%%%%%%%%%%%%%%%
\label{norm}%%%%%%%%%%%%%%%%%%%
%%%%%%%%%%%%%%%%%%%%%%%%%%%%%%%
\end{eqnarray}
where the summation is performed over the TQD atoms.

\begin{figure}[btp]
\centering
\leavevmode
\epsfxsize=8.5cm
\epsfbox{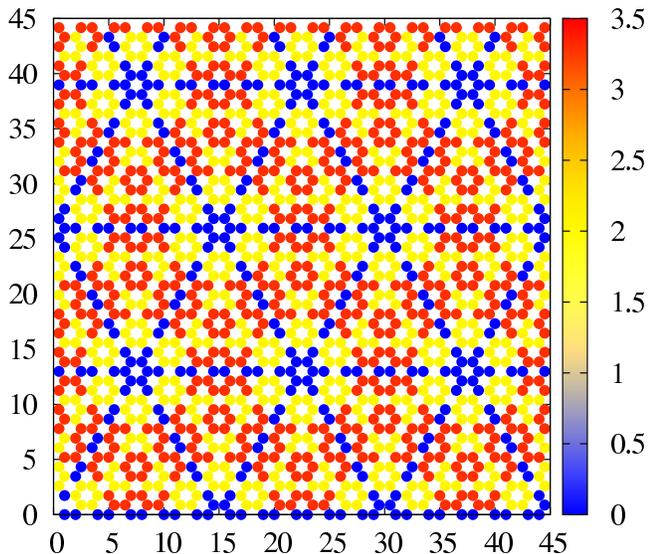}
\caption[]
{\label{latt}
(Color online.) The absolute value of the wave function
$\Psi^{n,m}$
on a large graphene lattice. The wave function vanishes on the black sites
(blue sites when the figure is in color), which are the auxiliary atoms.
The lines of the auxiliary atoms split the whole lattice in
$n_{\rm dot}$
triangular dots. Note that a given auxiliary atom is shared by two dots.
The bar on the right shows the correspondence between the dot color and the
wave function value, with blue equal to zero and red equal to 3.5.
}
\end{figure}

To find 
${\gamma}$ 
we use the following trick. Let us now consider a large lattice 
${\cal L}$,
whose linear size is much larger than $3N$, the size of our TQD. Consider,
further, a spinor wave function
$\Psi^{n,m}$
on such a lattice (see
Fig.~\ref{latt}).
This wave function vanishes on certain sites of the lattice, splitting the
whole area into
$n_{\rm dot}$ 
triangular dots. Clearly, the sites where the wave function vanishes
correspond to the auxiliary atoms.
%%%%%%%%%%%%%%%%%%%%%%%%%%%%%%%%%%%%%%%%%%%
%%%%%%%%%%%%%%%%%%%%%%%%%%%%%%%%%%%%%%%%%%%

Using the methods of
subsection~\ref{symmetry}
it is possible to prove that the wave functions on any two TQD of
Fig.~\ref{latt}
are connected by a unitary transformation. Therefore, the summation in
Eq.~(\ref{norm}),
performed over any TQD of
Fig.~\ref{latt},
gives unity. Thus, the summation over the entire lattice
${\cal L}$
gives us the number of the dots:
\begin{eqnarray}
{\gamma}^2
\left(
	\sum_{{\bf R} \in {\cal L}}
		|\psi^{\cal A}_{\bf R}|^2 
		+
	\sum_{{\bf R}+{\bm \delta}_1 \in {\cal L}}
		|\psi^{\cal B}_{{\bf R} + {\bm \delta}_1}|^2  
\right)
=
n_{\rm dot}.
\end{eqnarray}
On the other hand, the expression in the round brackets is equal to 
$6N_{\cal L}$,
where 
$N_{\cal L}$
is the number of atoms in 
${\cal L}$.
The factor of 6 appears because our wave function is composed of six
different plane waves. This is the advantage of introducing a large
lattice: we know that, when translational invariance is restored, the
interference between different plane waves of the sextet is negligible;
therefore, each plane wave contributes individually to the wave function
norm, and no cross-term needs to be calculated. Thus:
\begin{eqnarray}
6 {\gamma}^2 N_{\cal L} = n_{\rm dot}.
\end{eqnarray}
There are 
$N_a$
physical atoms 
and
$3N$
auxiliary atoms per TQD on 
${\cal L}$
(there are 
$6N$
auxiliary atoms surrounding one TQD, yet this amount has to be divided by
two, since any auxiliary atom is shared by two adjacent dots). In total,
there are $3N^2$ lattice sites per TQD. Therefore, we have
\begin{eqnarray}
n_{\rm dot} = \frac{N_{\cal L}}{3N^2}.
\end{eqnarray}
Combining the last two equations we derive:
\begin{eqnarray}
{\gamma} = \frac{1}{3\sqrt{2} N}.
\end{eqnarray}

\subsection{Single-electron state labelling}
\label{labelling}

It appears that for a pair of integer numbers, $n$ and $m$, there is a
unique single-electron state. This statement is incorrect: not every choice
of $n$, $m$ is allowed (for example, if
$n=m=0$,
then the corresponding wave function is exactly zero), and not every wave
function is unique (for example, if we rotate
${\bf k}_1$
by 
$2\pi/3$
we recover the same state).

It is necessary to introduce a scheme that uniquely labels every and any
quantum state. The most natural way of devising such a scheme is to
describe the allowed values of
${\bf k}^{n,m}$,
or, equivalently, of $n$ and $m$.

Specifying the allowed
${\bf k}^{n,m}$,
it quickly becomes obvious that the symmetric properties of the sextet are
important. Therefore, it is convenient to define the sextet's symmetry
group
$\tilde G$.
It is isomorphic to 
$C_{3v}$:
it consists of 
$\pm 2 \pi/3$
rotations around the origin and reflections about lines
$k_y = 0$,
$k_y = \pm \sqrt{3} k_x$.
Although, 
$\tilde G$
is isomorphic to the TQD's geometrical symmetry group $G$, they are not
identical: the reflection axes of $G$ do not coincide with those of 
$\tilde G$.

Developing this labelling system, one has to abide by the following
restrictions: 
(i) if
${\bf k}^{n',m'} = U {\bf k}^{n,m}$,
where 
$U \in \tilde G$,
then there exists a real number $\phi$, such that
$\Psi^{n', m'} = e^{i\phi} \Psi^{n, m}$;
(ii)
${\bf k}^{n,m}$
must lie within the graphene Brillouin zone;
(iii) any vector
${\bf k}^{n,m}$,
such that
$k_y^{n,m}=0$, or $k_y^{n,m} = \pm \sqrt{3} k_x^{n,m}$,
is disallowed: in this case the corresponding wave function vanishes
identically [see discussion after 
Eq.~(\ref{wf_vanish})];
(iv) there is no state when
${\bf k}^{n,m} = 0$
and when ${\bf k}^{n,m}$ is the location of the Dirac cone's apex;
(v) if
\begin{eqnarray}
U{\bf k}^{n',m'} - {\bf k}^{n,m} = {\bf d},\ U \in \tilde G,
%%%%%%%%%%%%%%%%%%%%%%%%%%%%%%%%%%
\label{(v)}%%%%%%%%%%%%%%%%%%%%%%%
%%%%%%%%%%%%%%%%%%%%%%%%%%%%%%%%%%
\end{eqnarray} 
where 
${\bf d}$ 
is the reciprocal lattice vector, then
$\Psi^{n', m'} = e^{i\phi} \Psi^{n, m}$.

\begin{figure}[btp]
\centering
\leavevmode
\epsfxsize=8.5cm
\epsfbox{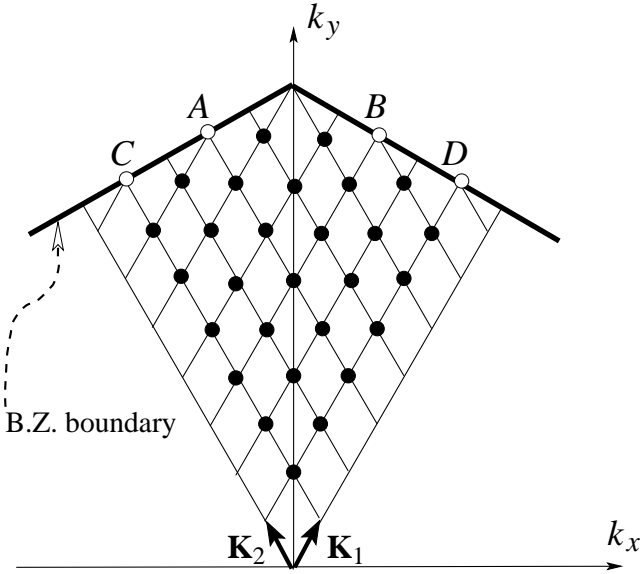}
\caption[]
{\label{label}
The allowed values of ${\bf k}^{n,m}$ occupy the sector 
$\sqrt{3}|k_x| < k_y$
of the graphene Brillouin zone (this sector is drawn in white in
Fig.~\ref{bz}).
Every filled circle represents a state. Points $A$ and $B$ (open
circles near the top) correspond to the same state. This is also true for
$C$ and $D$, see
subsection~\ref{labelling}. 
The thick solid line at the top of the figure is the Brillouin zone boundary.
}
\end{figure}

Keeping these conditions in mind let us consider the
following values for $n$ and $m$:
\begin{eqnarray}
n\geq 1, 
%%%%%%%%%%%%%%%%%%%%%%%%%%%%%%%%%%
\label{allow_n}%%%%%%%%%%%%%%%%%%%
%%%%%%%%%%%%%%%%%%%%%%%%%%%%%%%%%%
\\
m \geq 1,
%%%%%%%%%%%%%%%%%%%%%%%%%%%%%%%%%%
\label{allow_m}%%%%%%%%%%%%%%%%%%%
%%%%%%%%%%%%%%%%%%%%%%%%%%%%%%%%%%
\\
{\bf k}^{n,m} \in {\rm B.Z.}, 
%%%%%%%%%%%%%%%%%%%%%%%%%%%%%%%%%%%
\label{allow_k}%%%%%%%%%%%%%%%%%%%%
%%%%%%%%%%%%%%%%%%%%%%%%%%%%%%%%%%%
\\
{\bf k}^{n,m} \ne (0, 4\pi/(3\sqrt{3} a_0)).
%%%%%%%%%%%%%%%%%%%%%%%%%%%%%%%%%%%
\label{no_K}%%%%%%%%%%%%%%%%%%%%%%%
%%%%%%%%%%%%%%%%%%%%%%%%%%%%%%%%%%%
\end{eqnarray}
Here `B.Z.' stands for `Brillouin zone'. The allowed vectors 
${\bf k}^{n,m}$
lie within the white polygon of 
Fig.~\ref{bz}.
For
$N=5$
these vectors are shown in
Fig.~\ref{label}.
%%%%%%%%%%%%%%%%%%%%%%%%%%%%%%%%%%%%%%%%%%%%%%%%%

Observe that the condition (i) is met: indeed,
any two allowed vectors,
${\bf k}^{n,m}$
and
${\bf k}^{n',m'}$,
${\bf k}^{n',m'} \ne {\bf k}^{n,m}$,
cannot be connected by
$\tilde G$
transformations. Conditions (ii)-(iv) are explicitly satisfied. 

As for condition (v), it is necessary to realize that it is relevant only
if
${\bf k}^{n,m}$,
${\bf k}^{n',m'}$
lie on the zone's boundary. Otherwise, either
${\bf k}^{n,m}$,
or
${\bf k}^{n',m'}$
is outside of the zone. One can demonstrate that, to satisfy
Eq.~(\ref{(v)}),
the equality
$k_x^{n,m} = -k_x^{n',m'}$
must hold. Thus, in
Fig.~\ref{label},
points $A$ and $B$ (open circles) correspond to the identical state. The
same is true about $C$ and $D$.

%%%%%%%%%%%%%%%%%%%%%%%%%%%%%%%%%%%%%%%%%%%%%%%%%%%%%%%%%%%%
\begin{figure}[btp]
\centering
\leavevmode
\epsfxsize=8.5cm
\epsfbox{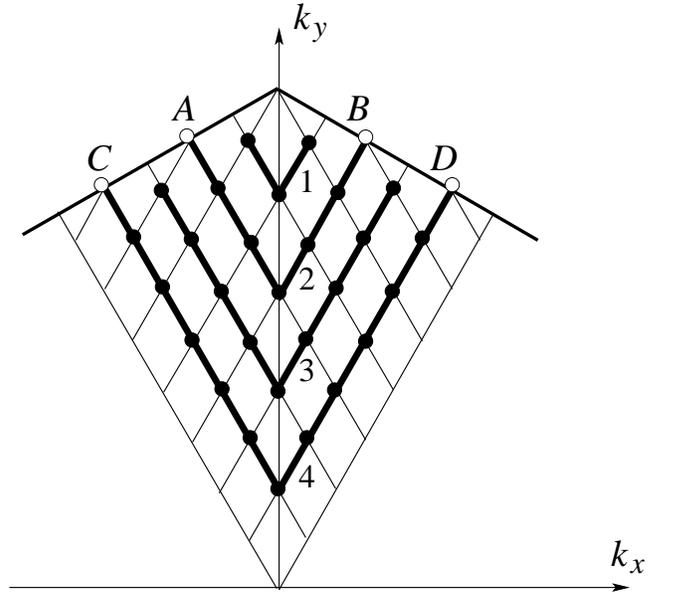}
\caption[]
{\label{count}
Counting the quantum states of a TQD. The thick solid V-shaped lines show
how we group our states to form an arithmetic progression. Numbers from 1
to 4 enumerate terms of the progression. Note that each pair of open
circles (points $A$ and $B$; $C$ and $D$) counts as one state, see
subsection~\ref{labelling}.
}
\end{figure}
%%%%%%%%%%%%%%%%%%%%%%%%%%%%%%%%%%%%%%%%%%%%%%%%%%%%%%%%%%%%

Finally, we want to count the total number of allowed states. It is
convenient to group the TQD states as shown in Fig.~\ref{count}. That way
they form an arithmetic progression: 3 states in the first group, 6 states
in the second group (five states inside the Brillouin zone and one state at
the zone's boundary), 9 states in the third group, etc. There are
$(N-1)$
terms in this progression. The sum of all terms, from the first to the
$(N-1)$th
is equal to:
\begin{eqnarray}
\frac{3}{2} N (N-1).
\end{eqnarray}
Since for every
${\bf k}^{n,m}$
there are two states,
$\Psi_{{\bf R}+}^{n,m}$ 
and
$\Psi_{{\bf R}-}^{n,m}$,
the above value has to be doubled. Therefore, the total number of states is
equal to 
\begin{eqnarray}
N_{\rm states} = 3 N (N-1).
\end{eqnarray}
We can see that 
$N_{\rm states} = N_a$.
This means that our labelling scheme is exhaustive; that is, there are no
states unaccounted by it.

\section{Corrections due to edge bond deformations}
\label{deformations}

In this section we apply the solution of the Schr\"odinger equation for a 
TQD to calculate the correction to the single-electron levels due to the
deformation of the carbon-carbon bonds at the edges of the TQD.

The edge bonds deformation is known to appear at the edges of graphene
nanoribbons \cite{fujita,louieI,gunlycke}. The deformation is not specific
to nanoribbons. Rather, it is a response of a carbon-carbon bond to an
atypical location (in this case, at the edge versus bulk). Thus, it is
likely that such deformation would be present at the edges on a TQD, should
this device be realized experimentally.

Our previous calculations completely disregard the edge deformation.
Fortunately, since the deformation is weak and since the number of deformed
bonds is much smaller than the number of undeformed bonds in a sufficiently
large TQD, such modification of the original problem can be accounted
within the framework of perturbation theory. Below we show how the
deformation of the edge bonds affects the single-electron eigenenergies.

At the Hamiltonian level, we now assume that the hopping amplitude across the
deformed bond
$t_{\rm d}$ 
deviates from $t$ \cite{louieI}:
\begin{eqnarray}
t_{\rm d} = t + \delta\! t.
\end{eqnarray}
The locations of the deformed bonds are shown in Fig.~\ref{tqd} by thick
solid lines. 

The Hamiltonian due to edge deformations is:
\begin{eqnarray}
\delta\! H
=
\delta\! H_{\rm lower \ edge} +
\delta\! H_{\rm left \ edge} +
\delta\! H_{\rm right \ edge},
\end{eqnarray}
where the three terms on the right-hand side of the equation correspond to
the three edges of the dot.

Let us first discuss the effect due to 
$\delta\! H_{\rm lower \ edge}$.
The deformed bonds at the lower edge connect two atoms within the same
primitive cell. These cells' positions are [see Eq.~(\ref{subA})]:
\begin{eqnarray}
{\bf R}_l = a_0 (1/2 + 3l, \sqrt{3}/2),
\\
1 \leq l \leq N-1,
\quad l - {\rm integer}.
\end{eqnarray} 
The matrix element between two arbitrary states 
$\Psi_{{\bf R}+}$
and
$\Phi_{{\bf R}+}$
is equal to:
\begin{eqnarray}
M
=
\langle \Phi_+|
\delta\! H_{\rm lower\! \ edge}
|\Psi_+ \rangle
= - \delta\! t \sum_{l=1}^{N-1}
			\Phi_{{\bf R}_l +}^*
		 	\sigma_x
			\Psi_{{\bf R}_l +}.
\end{eqnarray}
When 
$\Phi_{{\bf R} +}=\Psi_{{\bf R} +}$,
the matrix element is:
\begin{eqnarray}
M =
\delta\! t {\gamma}^2
\sum_{\alpha, \alpha' = 1}^6
	(-1)^{\alpha + \alpha'} 
	\left[
		\exp(-i\theta_{{\bf k}_\alpha})
		+
		\exp(i \theta_{{\bf k}_{\alpha'}})
	\right]
\quad
\\
\nonumber
\times
\sum_{l=1}^{N-1}
	\exp[ - i (
			{\bf k}_\alpha 
		   	- 
		   	{\bf k}_{\alpha'}
		  )
		  {\bf R}_l
	    ].
\quad
\end{eqnarray}
We can evaluate the sum over $l$:
\begin{eqnarray}
\sum_{l=1}^{N-1}
	\exp[ - i (
			{\bf k}_\alpha 
		   	- 
		   	{\bf k}_{\alpha'}
		  )
		  {\bf R}_l
	    ]
\quad
\quad
\\
\nonumber 
=
\exp\left[
	-\frac{i}{2}( k_{x \alpha} - k_{x \alpha'})a_0
	-\frac{\sqrt{3}i}{2} ( k_{y \alpha} - k_{y \alpha'})a_0
    \right]
\\
\nonumber
\times
\sum_{l=1}^{N-1}
	\exp[
		-3 i( k_{x \alpha} - k_{x \alpha'})a_0 l
            ].
\quad
\quad
\end{eqnarray}
The sum of the geometric series:
\begin{eqnarray}
\sum_{l=1}^{N-1}
	\exp[
		-3 i( k_{x \alpha} - k_{x \alpha'})a_0 l
            ]
\quad
\quad
\nonumber 
\\
\nonumber
=
\frac{
 	\exp[ - 3iN(k_{x \alpha} - k_{x \alpha'}) a_0 ]
	-
	\exp[ - 3i(k_{x \alpha} - k_{x \alpha'}) a_0 ]
     }
     {
	\exp[ - 3i(k_{x \alpha} - k_{x \alpha'}) a_0 ] - 1
     }.
\\
\end{eqnarray}
Depending on $\alpha$ and $\alpha'$, the quantity
$(k_{x \alpha} - k_{x \alpha'})$
is equal to:
\begin{eqnarray}
k_{x \alpha} - k_{x \alpha'} 
=
\cases{
	0,\cr 
	\pm \left(
			\frac{3}{2} k_x
			\pm
			\frac{\sqrt{3}}{2} k_y
	    \right), \cr
	\pm \sqrt{3} k_y.}
\end{eqnarray}
Thus, with the help of the condition
Eq.~(\ref{quant_cond_vec})
we can write:
\begin{eqnarray}
3N(k_{x \alpha} - k_{x \alpha'}) a_0
=
\cases{
	0, \cr
	\pm \pi (n-m) \pm \pi (n+m), \cr
	\pm 2 \pi (n+m).
}
\end{eqnarray}
Therefore, for any $\alpha$ and $\alpha'$ it holds that:
\begin{eqnarray}
\exp[-3iN(k_{x \alpha} - k_{x \alpha'}) a_0] = 1 
\; \Rightarrow
\\
\sum_{l=1}^{N-1}
	\exp[
		-3 i( k_{x \alpha} - k_{x \alpha'})a_0 l
            ]
\\
\nonumber
=
\cases{
	-1, & for\ \ \ $(k_{x \alpha} - k_{x \alpha'}) \ne 0$,\cr
	N-1, & for\ \ \ $(k_{x \alpha} - k_{x \alpha'}) = 0$.
}
\end{eqnarray}
Expressing the last formula differently, one writes:
\begin{eqnarray}
\sum_{l=1}^{N-1}
	\exp[
		-3 i( k_{x \alpha} - k_{x \alpha'})a_0 l
            ]
=
-1 + N \delta_{k_{x \alpha}, k_{x \alpha'}}.
%%%%%%%%%%%%%%%%%%%%%%%%%%%%%%%%%%%%%%%%%%
\label{summ_geom}%%%%%%%%%%%%%%%%%%%%%%%%%
%%%%%%%%%%%%%%%%%%%%%%%%%%%%%%%%%%%%%%%%%%
\end{eqnarray} 
The matrix element $M$ can be written as a sum:
\begin{eqnarray}
M = M_0 + N M_1,
%%%%%%%%%%%%%%%%%%%%%%%%%%%%%%%%%%%%%%%%%%
\label{M+M}%%%%%%%%%%%%%%%%%%%%%%%%%%%%%%%
%%%%%%%%%%%%%%%%%%%%%%%%%%%%%%%%%%%%%%%%%%
\end{eqnarray}
where the $M_0$ term corresponds to $-1$ in the right-hand side of
Eq.~(\ref{summ_geom}),
and 
$N M_1$ 
terms corresponds to the Kronecker delta there. For
$M_1$
we obtain:
\begin{eqnarray}
M_1
=
\delta\! t {\gamma}^2
\sum_{\alpha = 1}^6
	2 \cos \theta_{{\bf k}_\alpha}
	[ 1 - \cos (\sqrt{3} k_{y \alpha} a_0) ].
%%%%%%%%%%%%%%%%%%%%%%%%%%%%%%%%%%%%
\label{M1}%%%%%%%%%%%%%%%%%%%%%%%%%%
%%%%%%%%%%%%%%%%%%%%%%%%%%%%%%%%%%%%
\end{eqnarray}
The first term in the brackets corresponds to summands for which $\alpha'$
is such that
$k_{x \alpha} = k_{x \alpha'}$
and
$k_{y \alpha} = k_{y \alpha'}$.
The second term corresponds to summands for which $\alpha'$ is such that
$k_{x \alpha} = k_{x \alpha'}$
and
$k_{y \alpha} = - k_{y \alpha'}$.

To evaluate the sum
$\sum_{\alpha } \cos \theta$,
it is convenient to use Eq.~(\ref{theta_def}) and
Eq.~(\ref{graphene_energy}):
\begin{eqnarray}
\sum_{\alpha = 1}^6
	2 \cos \theta_{{\bf k}_\alpha}
=
\sum_{\alpha = 1}^6
\frac{
	t^{\vphantom{*}}_{{\bf k}_\alpha}
	+
	t^*_{{\bf k}_\alpha}
     }
     {
	\varepsilon_{{\bf k}_\alpha +}
     }.
\end{eqnarray}
Since the energy is independent of the index $\alpha$:
$\varepsilon_{{\bf k}_\alpha +} = \varepsilon_{{\bf k}+}$,
one can write the following expression for this sum:
\begin{eqnarray}
\sum_{\alpha = 1}^6
	2 \cos \theta_{{\bf k}_\alpha}
\quad
\quad
%%%%%%%%%%%%%%%%%%%%%%%%%%%%%%%%%%%
\label{sum_1}%%%%%%%%%%%%%%%%%%%%%%
%%%%%%%%%%%%%%%%%%%%%%%%%%%%%%%%%%%
\\
\nonumber
=
\frac{4t}{\varepsilon_{{\bf k}+}}
\sum_{\alpha = 1, 3, 5}
	\left[
		1 + 2 \cos\left(
				\frac{3 k_{x \alpha} a_0}{2}
			  \right)
		      \cos\left(
				\frac{\sqrt{3} k_{y \alpha} a_0}{2}
			  \right)
	\right].
\end{eqnarray}
Substituting the formulas for 
${\bf k}_{1,3,5}$,
Eq.~(\ref{k1}), Eq.~(\ref{k3}), and Eq.~(\ref{k5}), into 
Eq.~(\ref{sum_1})
one obtains:
\begin{eqnarray}
\sum_{\alpha = 1}^6
	2 \cos \theta_{{\bf k}_\alpha}
=
\frac{4 \varepsilon_{{\bf k}+}}{t}.
\end{eqnarray}
The calculation of the second term of Eq.~(\ref{M1}) is performed along the
same lines. The result is:
\begin{eqnarray}
\sum_{\alpha = 1}^6
	2 \cos \theta_{{\bf k}_\alpha} \cos ( \sqrt{3} k_{y \alpha} a_0)
\\
\nonumber
=
\frac{4 \varepsilon_{{\bf k}+}}{t}
+
\frac{4 t}{\varepsilon_{{\bf k}+}}
\Big[
	\cos (3k_x a_0) 
\\
\nonumber 
	+ 
	2\cos \left(\frac{3}{2} k_x a_0 \right)
	\cos \left( \frac{3\sqrt{3}}{2} k_y a_0 \right) 
	- 
	3
\Big].
\end{eqnarray}
Therefore, we can express $M_1$ as follows:
\begin{eqnarray}
M_1 
&=&
-\frac{4 \gamma^2 t \delta\! t}{\varepsilon_{{\bf k}+}}
\Big[
	\cos (3k_x a_0 ) 
\\
\nonumber 
	&+&
	2\cos \left(\frac{3}{2} k_x a_0  \right)
	\cos \left( \frac{3\sqrt{3}}{2} k_y a_0  \right) 
	- 
	3
\Big]
\\
\nonumber
&=&
\frac{2 \gamma^2 t \delta\! t}{\varepsilon_{{\bf k}+}}
[6 - F( \sqrt{3} \tilde{{\bf k}})],
\end{eqnarray}
where 
$\tilde{\bf k} = (k_y, k_x)$,
and the function $F$ is defined by Eq.~(\ref{F}).

The evaluation of $M_0$ from Eq.~(\ref{M+M}) is easy to perform:
\begin{widetext}
\begin{eqnarray}
%%%%%%%%%%%%%%%%%%%%%%%%%%%%%
\label{M0}%%%%%%%%%%%%%%%%%%%
%%%%%%%%%%%%%%%%%%%%%%%%%%%%%
M_0 
=
-\delta\! t {\gamma}^2
\sum_{\alpha, \alpha'  = 1}^6
(-1)^{\alpha + \alpha'}
%\quad
%\\
%\nonumber
%\times
\exp\left[
	-\frac{i}{2}( k_{x \alpha} - k_{x \alpha'})a_0
	-\frac{\sqrt{3}i}{2} ( k_{y \alpha} - k_{y \alpha'})a_0
    \right]
%\\
%\nonumber
%\times
\left[
	\exp(-i \theta_{{\bf k}_\alpha})
	+
	\exp(i \theta_{{\bf k}_{\alpha'}})
\right]
\\
\nonumber
=
-\delta\! t {\gamma}^2
\sum_{\alpha = 1}^6
(-1)^{\alpha }
%\\
%\nonumber 
%\times
\exp\left(
	-\frac{i}{2} k_{x \alpha} a_0
	-\frac{\sqrt{3}i}{2}  k_{y \alpha} a_0
    \right)
\exp(-i \theta_{{\bf k}_\alpha})
%\\
%\nonumber
%\times
\sum_{\alpha' = 1}^6
(-1)^{\alpha'}
\exp\left(
	\frac{i}{2} k_{x \alpha'}a_0
	+
	\frac{\sqrt{3}i}{2} k_{y \alpha'} a_0
    \right)
+ {\rm C.c.},
\end{eqnarray}
\end{widetext}
where `C.c.' stands for the complex-conjugated terms.

The sum over $\alpha'$ in Eq.~(\ref{M0}) is zero. To prove this let us
rewrite it:
\begin{eqnarray}
\sum_{\alpha' = 1}^6
(-1)^{\alpha'}
\exp\left(
	\frac{i}{2} k_{x \alpha'}a_0
	+
	\frac{\sqrt{3}i}{2} k_{y \alpha'} a_0
    \right)
%%%%%%%%%%%%%%%%%%%%%%%%%%%%%%%%%%%%%%
\label{sum}%%%%%%%%%%%%%%%%%%%%%%%%%%%
%%%%%%%%%%%%%%%%%%%%%%%%%%%%%%%%%%%%%%
\\
\nonumber
=
\sum_{\alpha' = 2,4,6}
\exp ( i a_0 {\bf u} {\bf k}_{\alpha'})
-
\sum_{\alpha' = 1,3,5}
\exp ( i a_0 {\bf u} {\bf k}_{\alpha'}).
\end{eqnarray}
Examining Fig.~\ref{sextet} it becomes obvious that
${\bf u} {\bf k}_1={\bf u} {\bf k}_6$,
${\bf u} {\bf k}_3={\bf u} {\bf k}_4$,
and
${\bf u} {\bf k}_2={\bf u} {\bf k}_5$.
This implies that both terms on the right-hand side of Eq.~(\ref{sum}) are
equal, and they cancel each other exactly.

Combining the above results, we write for $M$:
\begin{eqnarray}
M
=
\frac{2 \gamma^2 N t \delta\! t}{\varepsilon_{{\bf k}+}}
[6 - F( \sqrt{3} \tilde{{\bf k}})].
\end{eqnarray}
This expression gives the matrix element for the operator corresponding to
the bond deformations at the lower edge of the TQD. The matrix element for
the bond deformations at all three edges is equal to $3M$:
\begin{eqnarray}
\delta\! H^{n,m}_+
=
\langle \Psi_+^{n,m} |
\delta\! H
|\Psi_+^{n,m}  \rangle
=
\frac{t \delta\! t}{3 N \varepsilon_{{\bf k}+}}
[6 - F( \sqrt{3} \tilde{{\bf k}})].
\quad
\end{eqnarray}
The formula above can be generalized:
\begin{eqnarray}
\delta\! H^{n,m}_\pm
=
\langle \Psi_\pm^{n,m} |
\delta\! H
|\Psi_\pm^{n,m}  \rangle
=
\pm
\frac{\delta\! t}{3N}
\frac{6 - F( \sqrt{3} \tilde{{\bf k}})}{ \sqrt{ 3 + F({\bf k})}},
%%%%%%%%%%%%%%%%%%%%%%%%%%%%%%%%%%%%
\label{matrix_elm}%%%%%%%%%%%%%%%%%%
%%%%%%%%%%%%%%%%%%%%%%%%%%%%%%%%%%%%
\end{eqnarray}
to account for the states with negative energies.

To evaluate first-order corrections to the eigenenergies due to 
$\delta\! H$
it is necessary to find not only the diagonal elements 
$\delta\! H^{n,m}_\pm$,
but the off-diagonal elements, connecting the degenerate states, as well.
In our case, two wave functions
$\Psi^{n,m}$
and
$\Psi^{m,n}$
correspond to degenerate states, unless 
$n = m$,
or the vector
${\bf k}^{n,m}$
lies on the Brillouin zone boundary. However, the element
$\langle \Psi^{n,m}| \delta\! H | \Psi^{m,n} \rangle$ 
vanishes. Indeed,
$\delta\! H$ is invariant under transformations from 
$G$
and, therefore, the matrix element is non-zero only if both states transform
identically under 
$G$.
The latter condition is never fulfilled, for degenerate wave functions
either acquire different phase factors upon the rotations,
Eq.~(\ref{angular_mom}), or they transform differently when subjected to
reflections, Eq.~(\ref{A12}) and Eq.~(\ref{refl}).

The above considerations show that the correction to the eigenenergies are
given by Eq.~(\ref{matrix_elm}). Let us discuss this expression.

First of all, we notice that
$[6 - F( \sqrt{3} \tilde{{\bf k}})]/ \sqrt{ 3 + F({\bf k})}$
is positive. Therefore, the sign of the correction is determined by the
sign of
$\pm \delta\! t $.

Second, since $F$ is an even function of its arguments, the degeneracy
between 
$\Psi^{n,m}$
and
$\Psi^{m,n}$
remains.

Third, the larger the dot, the smaller the correction: the characteristic
energy scale for the correction is 
$\delta\! t/N$,
which decreases when $N$ grows.  This is natural, since the ratio of the
deformed bonds 
($\sim N$)
to the total number of bonds in a TQD
($\sim N^2$)
decreases when the dot increases.

\section{Discussion}
\label{discussion}

In this paper we find the exact spectrum of a graphene TQD with armchair
edges.
Certain matrix elements are evaluated with the help of our wave functions.
Thus, our solution may be used for perturbation theory calculations, e.g.,
for weak magnetic field, disorder. 

\begin{figure}[btp]
\centering
\leavevmode
\epsfxsize=8.5cm
\epsfbox{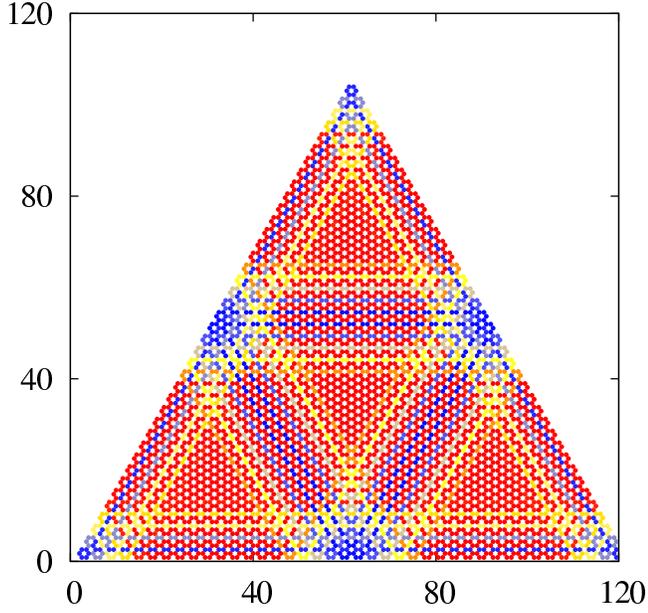}
\caption[]
{\label{state_a}
(Color online.) Probability density for the state with
$n=39$
and
$m=41$
for a triangular graphene dot. The dot's effective size,
$L = 3Na_0$,
is fixed by the value of the constant
$N=41$.
The total number of atoms in such a dot is
$N_a = 4920$.
The probability density for the same state is presented in Fig.~4(a) of
Ref.~\cite{akola_PRB}.
}
\end{figure}

\begin{figure}[btp]
\centering
\leavevmode
\epsfxsize=8.5cm
\epsfbox{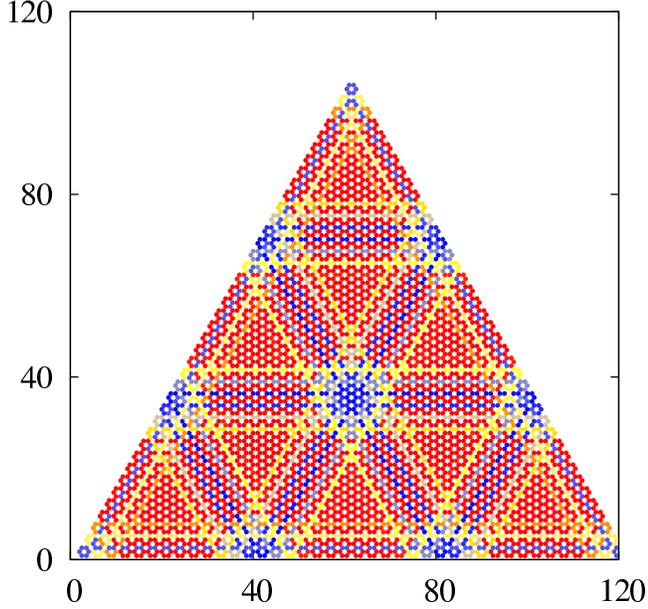}
\caption[]
{\label{state_c}
(Color online.) Probability density for the state with
$n=38$
and
$m=41$
for a triangular graphene dot. The dot's effective size,
$L = 3Na_0$,
is fixed by the value of constant
$N=41$.
The total number of atoms in such a dot is
$N_a = 4920$.
The probability density for the same state is presented in Fig.~4(c) of
Ref.~\cite{akola_PRB}.
}
\end{figure}

The problem of the electronic properties of graphene TQD is addressed
numerically in several papers (e.g.,
\cite{qd_numerics,akola,akola_PRB}).
To show that our analytical approach agrees with numerical solutions, we
calculated the probability density for the states with
(a) 
$n=39$,
$m=41$
(Fig.~\ref{state_a}),
and (b) 
$n=38$,
$m=41$
(Fig.~\ref{state_c}),
both for a TQD with
$N=41$.
There are 
$N_{\rm a} = 4920$
atoms in such a TQD. The eigenenergies of states (a) and (b) are close to
zero. 
%%%%%%%%%%%%%%%%%%%%%%%%%%%%%%%%%%%%%%%%%%%
%%%%%%%%%%%%%%%%%%%%%%%%%%%%%%%%%%%%%%%%%%%
%%%%%%%%%%%%%%%%%%%%%%%%%%%%%%%%%%%%%%%%%%%
%%%%%%%%%%%%%%%%%%%%%%%%%%%%%%%%%%%%%%%%%%%

These states are chosen here because their probability distributions are
mapped in
Ref.~\cite{akola_PRB}.
Comparing Fig.~4(a) of the latter reference and our
Fig.~\ref{state_a}
we see that the probability density distributions
are similar. The same is true about Fig.~4(c) of
Ref.~\cite{akola_PRB}
and our
Fig.~\ref{state_c}.

To conclude, generalizing the existing solution, we find the exact wave
functions and eigenenergies for an electron inside a graphene TQD. The
symmetry properties of our wave functions are determined. As an
application, the corrections to the eigenenergies due to the edge bonds'
deformations are calculated. We also demonstrate that our exact solution is
in agreement with previous numerical work.

\section{Acknowledgements}

We are grateful for the support provided by grant RFBR-JSPS 09-02-92114.
FN gratefully acknowledges partial support from the National Security
Agency
(NSA), Laboratory Physical Sciences (LPS), Army Research Office (ARO), and
National Science Foundation (NSF) grant No.~0726909.
%

%%%%%%%%%%%%%%%%%%%%%%----BIBLIOGRAPHY----%%%%%%%%%%%%%%%%%%%%%%%%%%%%%%%%%

\newpage

\end{document}